# Memristive properties and synaptic plasticity in substituted pyridinium iodobismuthates

Gisya Abdi, *[a] Tomasz Mazur,[a] Ewelina Kowalewska,[a] Andrzej Sławek,[a] Mateusz Marzec,[a] Konrad Szaciłowski*[a,b]

[1]AGH University of Krakow, Academic Centre for Materials and Nanotechnology, al. Mickiewicza 30, 30-059 Kraków, Poland

[2]Unconventional Computing Lab, University of the West of England, Bristol BS16 1QY, United Kingdom

*corresponding authors: agisya@agh.edu.pl, szacilow@agh.edu.pl, konrad.szacilowski@uwe.ac.uk

**Abstract:** This study explores the impact of organic cations in bismuth iodide complexes on their memristive behavior in metal-insulator-metal (MIM) type thin-layer devices. The presence of electron-donating and electron withdrawing functional groups (-CN, -CH$_3$, -NH$_2$, and -N(CH$_3$)$_2$) on pyridinium cations induces morphological alterations in crystals, thus influencing the electronic or ionic conductivity of devices comprising sandwiched thin layers (thickness = 200 nm ±50) between glass/ITO as bottom electrode (~ 110 nm) and copper (~ 80 nm) as the top electrode. It was found that the current-voltage (I-V) scans of the devices reveal characteristic pinched hysteresis loops, a distinct signature of memristors. The working voltage windows are significantly influenced by both the types of cation and the dimensionality of ionic fragments (0D or 1D) in the solid-state form. Additionally, the temperature alters the surface area of the I-V loops by affecting resistive switching mechanisms, corresponding log-log plots at three temperatures (-30 °C, room temperature and 150 °C) are fully studied. Given that a memristor can operate as a single synapse without the need for programming, aligning with the requirements of neuromorphic computing, the study investigates long-term depression, potentiation, and spike-time-dependent plasticity-a specific form of the Hebbian learning rule-to mimic biologically synaptic plasticity. Different polar pulses, such as triangle, sawtooth, and square waveforms were employed to generate Hebbian learning rules. The research demonstrates how the shape of the applied pulse series, achieved by overlapping pre- and post-pulses at different time scales, in association with the composition and dimensionality of ionic fragments, lead to changes in the synaptic weight percentages of the devices.

**Introduction**

The von Neumann concept, which segregates the data processing unit from data storage, underpins the architecture of the majority modern computer designs. The bottleneck known as the von Neumann bottleneck arises from the limited bandwidth in communication between memory and processors, significantly impacting overall computing performance. Neuromorphic architectures are emerging as promising alternatives to current computing systems due to their rapid operational speed, scalability options, and, crucially, low energy consumption.[1, 2] Although state-of-the-art neuromorphic complementary metal-oxide-semiconductor (CMOS) devices achieve remarkable energy efficiency, their significant practical application stems from incorporating both p- and n-type metal-oxide-semiconductor (MOS) components.[3, 4] However, the extensive evolution of CMOS transistor technology, driven by Moore's law, encounters challenges as devices approach atomic scales. Issues such as quantum effects, heat dissipation, and escalating manufacturing costs become significant. Power consumption, material limitations, and economic considerations further pose obstacles, prompting the exploration of alternative technologies and innovative design approaches to sustain the advancement. On the contrary, memristive and other neuromorphic devices, offer a simpler implementation with easy fabrication, compact size, and high stacking density.[5-9] Importantly, the functional layer of the memristor is not restricted to semiconductor materials. The memristor, characterised by a two-terminal device with a metal/insulator/metal (MIM) structure, displays its behaviour across a wide range of materials.[6, 10-12] Memristors' resistive-switching properties can be realised in a number of materials, albeit with varying dynamic resistance ranges. Low-dimensional materials (0D, 1D),[13-15] lead halide perovskites based on lead halides (or lead-free),[15, 16] 2D materials (graphene, MXenes, boron nitride, transition metal chalcogenides, metal-organic frameworks),[17-20] polymers, biopolymers[21-24], nanocomposites[15] are some reported classes of applied materials in these kinds of devices. The resistive switching characteristic and range of conductive states depend on the key mechanism and are particularly linked to the array of the components and the operating environment. Researchers have identified several potential mechanisms, including electrochemical metallisation (ECM), valence change mechanism (VCM), Schottky barrier modulation and other interfacial phenomena, ferroelectric polarisation, electronic (or electrostatic) interactions, and thermochemical process.[25] Three first mechanisms have been seen variously in devices of the perovskite type. ECM is merely a filamentary type process; however, VCM can be seen in both filamentary-based and interface-based processes. In the case of filamentary processes, devices are not fully scalable, as the size of the electrode can affect the conductivity of the fabricated device and the ratio of high resistance state (HRS) to low resistance state (LRS), commonly referred to as the ON-OFF ratio.[26] Consideration of low activation energy for ion migration in lead-based perovskite materials, such as methylammonium lead halide (MAPbX$_3$), has been explored for the development of memristor-based synapses with low operating voltages. However, challenges of toxicity and the presence of iodine vacancies in perovskite films persist. These vacancies result from the loss of iodide ions due to the volatilisation of methyl ammonium iodide (MAI) during the annealing process of perovskite films, thereby significantly restricting their potential for extensive applications.[27]

As a less harmful substitute, various bismuth complexes with diverse dimensionalities have been documented incorporating Bi-halides characterised by different fragment sizes and compositions. This diversity is determined by the potential for sharing faces, vertices, or edges of octahedra within the crystalline structures. However, in analogous bismuth-based counterparts, the films persistently generate new iodine vacancies, even when previously existing vacancies are filled with an excess of iodide sources. This continuous generation of iodine vacancies is identified as a probable cause behind the inadequate long-term stability observed in the fabricated devices. To address this issue, researchers have explored the substitution of methyl ammonium (MA) with larger and more stable organic salts. For instance, Haque et al. employed phenylethylammonium iodide to passivate the iodine vacancies and address morphological imperfections in perovskite films.[27, 28]

This study introduces a number of stable, less-toxic bismuth halides based on pyridinium that can be used in memristive devices. We explored the fine-tuning of pyridinium iodobismuthates(III) (pyBiI$_3$) by adjusting the electronic structure of the cations while preserving their geometric integrity as much as possible. We chose a series of structurally similar pyridinium cations and prepared 4-aminopyridinium iodobismuthate (4-AmpyBiI$_3$), 4-methylpyridinium iodobismuthate (4-MepyBiI$_3$), 4-dimethylaminopyridine (4-DmapyBiI$_3$) and 4-cyanopyridinium iodobismuthate (4-CNpyBiI$_3$) which, despite their nearly identical geometry, exhibit significant differences in electronic properties, such as electron donor or acceptor characteristics and dipole moments.[29]

We address the most recent developments in spike-time-dependent plasticity (STDP)[30, 31] in memristor-based artificial synapses by changing and applying different shapes of bipolar pulses and investigate how they affect the synaptic weights in the applied devices in STDP experiment; results have been discussed in detail. In pyridinium-based bismuth complexes, the effect of functional groups (electron withdrawing or electron donating substitution) can vary the conducting mechanism and working voltages and produce various electronic characteristics.

## Results and discussion

### Characterization of thin layers on ITO glass

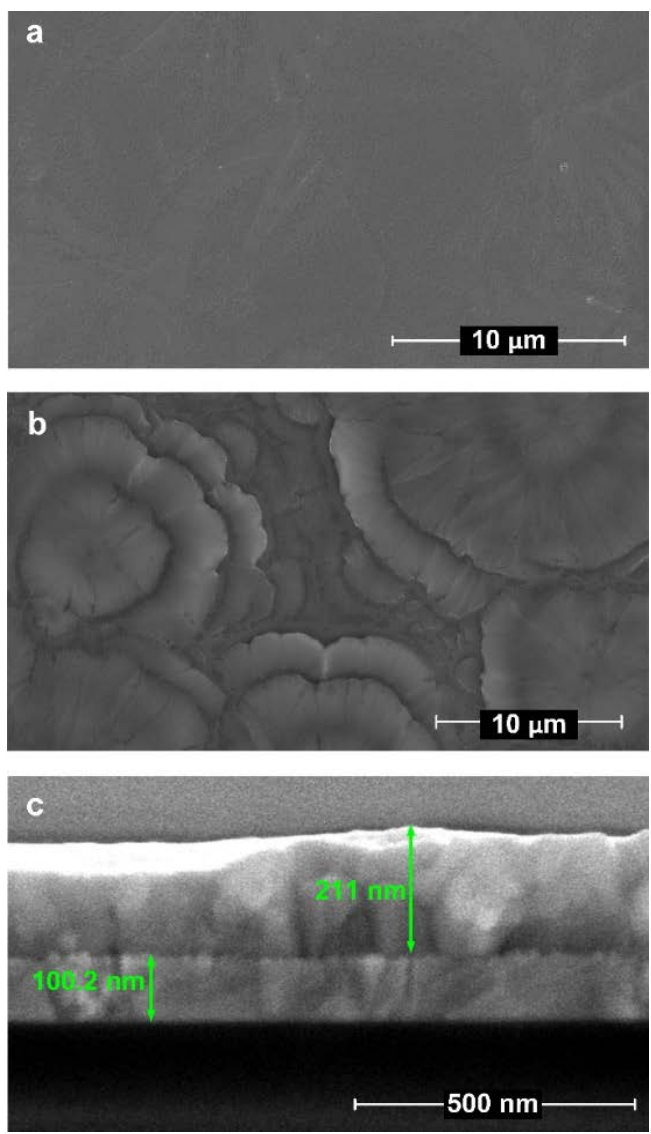

**Figure 1.** SEM images of 4-CNpyBiI$_3$: a) 300 mg / mL, b) 400 mg/mL c) cross-sectional SEM image of 4-CNpyBiI$_3$ /ITO/glass.

Thin layers of Bi complexes were prepared using two different concentrations: 300 mg/ml (solution 1) and 400 mg/mL (solution 2) of bismuth complexes dissolved in DMF. These layers were spin coated at 2000 rpm for 30 seconds, and the SEM results revealed distinct results for the two concentrations. In the case of the more concentrated solution 2, the thin layer exhibited nonuniformity, with crystals growing on the surface and protruding. In contrast, solution 1 produced a homogeneous and uniform surface (Figure 1), with layer thickness distributions ranging from approximately 150 to 220 nm, as confirmed by profilometry.

For 4-CNpyBiI$_3$, the cross-sectional SEM image displayed a uniform layer with a thickness of 211 nm (Figure 1). It should be noted that solutions with concentrations lower than those of solution 1 were excessively diluted, proving insufficient to cover the roughness of ITO and leading to potential short circuits in electrical measurements.

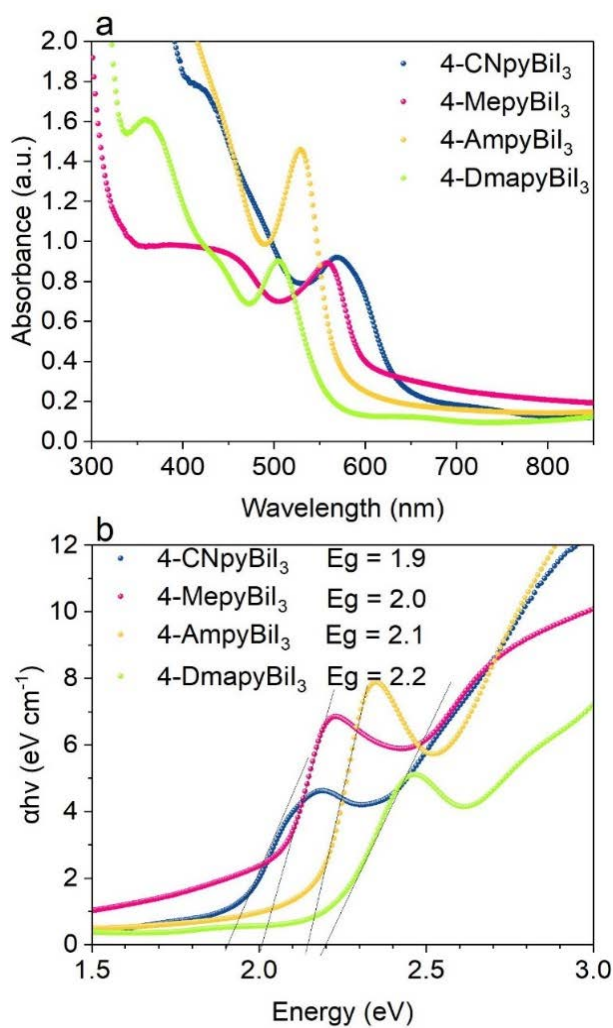

**Figure 2**. UV-vis absorption spectra (a) and Tauc plots (b) of thin layers of bismuth-based complexes on ITO glass.

Optical parameters of these layers were determined by means of UV-vis-NIR absorption spectroscopy and spectroscopic ellipsometry: band gap energies, real and imaginary dielectric functions, and layer thicknesses (Table S1).

Using the Tauc method, optical absorption spectroscopy was used to calculate the optical band gap of the manufactured devices. In this equation $(a.hv)^r = A(hv - E_g)$, where $h$ represents the Planck constant, $v$ denotes the photon frequency, $A$ is a constant, and $E_g$ stands for the band gap energy (Figure 2). The exponent $r$ fixed at 1 which is suitable for ionic/molecular crystals with only minor covalent component. The optical gap of pristine BiI$_3$ was estimated to be 1.6 eV, whereas the pyridinium-based samples exhibited significantly higher band gap values: $E_{g(4-CNpyBiI_3)} = 1.9$ eV, $E_{g(4-AmpyBiI_3)} = 2.1$, $E_{g(4-MepyBiI_3)} = 2.0$ eV and $E_{g(4-DmapyBiI_3)} = 2.2$ eV . The 4-CNpyBiI$_3$ withdrawing functional group shows the lowest band gap and 4-DmetpyBiI$_3$ with the electron donating group shows the highest band gap.

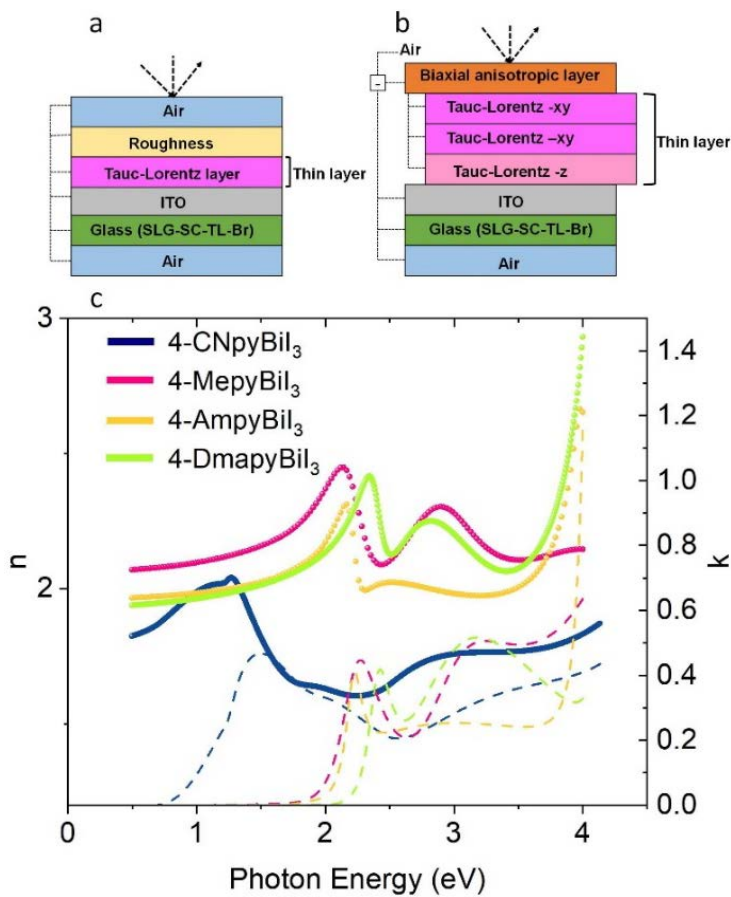

**Figure 3.** Schematic representation of the models used to fit experimental data from spectroscopic ellipsometry for (a) 4-MepyBiI$_3$, 4-DmapyBiI$_3$ and 4-AmpyBiI$_3$, (b) biaxial anisotropic model for 4-CNpyBiI$_3$

as Tauc-Lorentz layers, and (c) refractive indices of pyridinium-based bismuthate samples. The thickness of the glass was fixed at 1.1 mm.

Spectroscopic ellipsometry was employed to measure the thickness and refractive index, an essential optical property of semiconductors. Incident light angle was applied in the 50-70° range with 5° intervals, and the recorded amplitude ratio (Ψ) and phase difference (Δ) are presented in Figure 3. The Tauc-Lorentz (TL) optical model was chosen to achieve the best fit with experimental data for thin layers of pyridinium-based bismuthate samples as absorbers. The fitted plots and the resulting refractive index (n), indicating phase velocity, and extinction or attenuation coefficient (k) are depicted against photon energy in Figure 3. The fitting plots are represented in Figure S1 and values at 632.8 nm (1.96 eV) and room temperature are reported in Table S1. The refractive indices for these samples exceed 2, comparable to the value reported for $CH_3NH_3PbI_3$ (n = 2.61) by Löper et al.[32] However, in the case of 4-CNpyBiI$_3$, due to the anisotropic property of the surface, the fitting data in the TL model did not exhibit good agreement between experimental and theoretical data. The biaxial anisotropic model was selected with two Tauc-Lorentz layers in the xy plane with six oscillators and one Tauc-Lorentz layer in the z direction, including six oscillators. The results reveal remarkably lower values of n and higher values of k compared to the former derivatives with electron-donating groups on the para position in the pyridinium cation. Band gaps estimated on the basis of imaginary part of dielectric function ($E_{g(4-CNpyBiI_3)} = 2.06$ eV, $E_{g(4-AmpyBiI_3)} = 2.09$, $E_{g(4-MepyBiI_3)} = 2.02$ eV and $E_{g(4-DmapyBiI_3)} = 2.24$ eV.) are in a good agreement with the data obtained by Tauc method (vide supra).

The work function, which represents the energy difference between photon energy and the cut-off position of the secondary electron (SE), along with the hole injection barrier, determined by the variation between the substrate Fermi level and the highest occupied molecular orbital (HOMO) onset of the material, was measured using ultraviolet photoelectron spectroscopy (UPS). Each UPS spectrum underwent a correction process to eliminate emission features stemming from secondary line excitations of the He–I gas discharge. Adjustments were made to the relative intensities of satellite excitations in the secondary line subspectra to align them accurately with the measured UPS spectrum, taking into account the variation in intensities due to He discharge pressure. The subtraction process commenced incrementally, starting with the highest photon energy satellite. To mitigate the potential material degradation caused by exposure to high-flux radiation, measurement durations were minimised.

UPS measurements were used to further scrutinize the electronic structure of pyridinium-based complexes, unveiling their classification as p-type semiconductors, indicating defects in the cationic sublattice (e.g. proton vacancies, compensated by excess of holes). Secondary cutoff energies were identified as 16.4, 16.3, 16.2, and 17.85 for 4-CNpyBiI$_3$ (Figure 4), 4-AmpyBiI$_3$, 4-MepyBiI$_3$ and 4-DmapyBiI$_3$ respectively (Figure S2). Additionally, the UPS results clearly demonstrate alterations in the electronic structures of bismuth-based complexes, suggesting potential dimensional changes from 0D to 1D that can impact key electronic parameters. These changes include variations in the hole injection barrier ($\Phi_B$), ionisation energy (IE), and work function of p-type semiconductors ($\Phi_S$), as summarised in Table 1. The work function holds significance in guiding the selection of an appropriate metallic electrode for sputtering, crucial for predicting the behaviour of resulting metal/p-type semiconductor contacts.

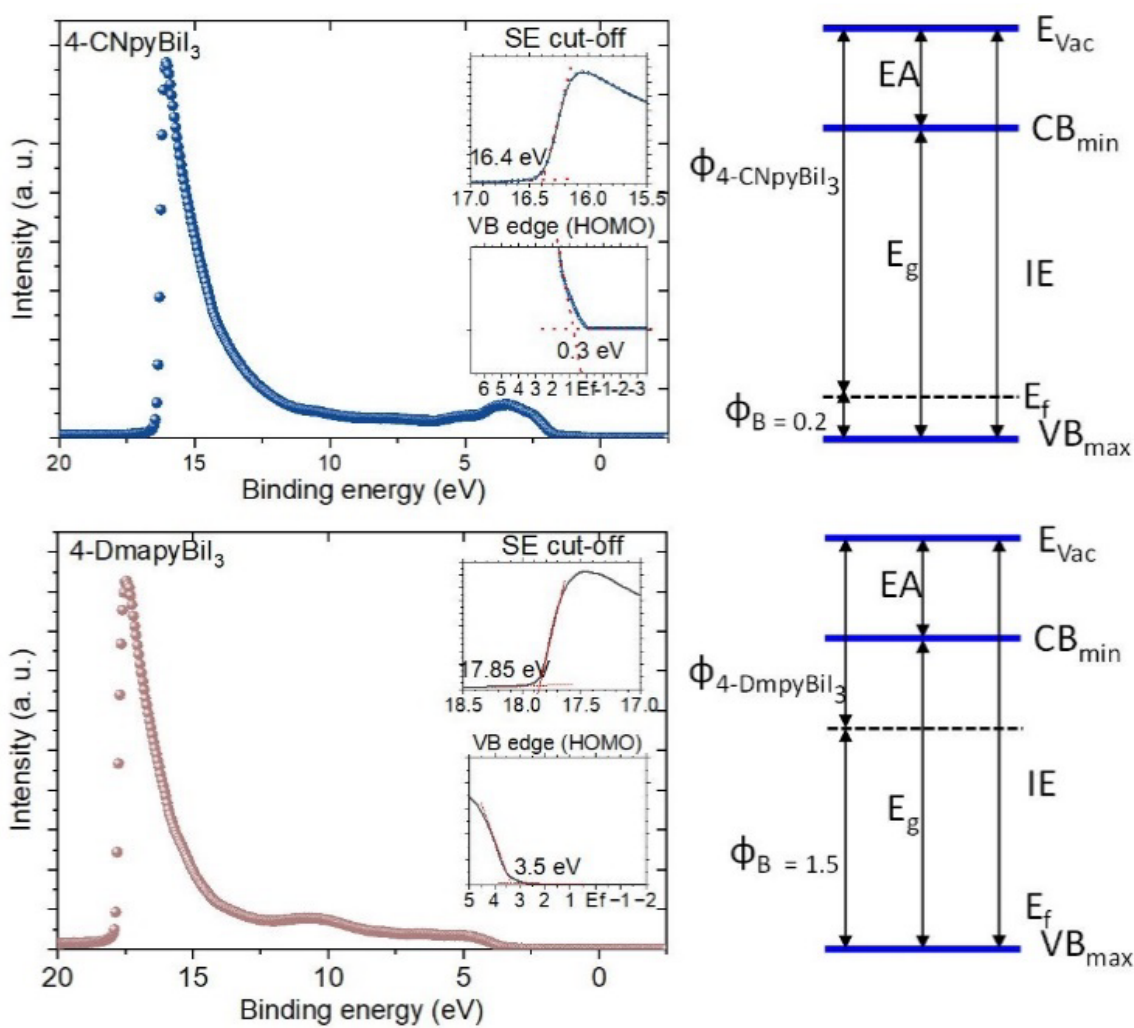

**Figure 4.** UPS spectra of the 4-CNpyBiI$_3$ and 4-DmapyBiI$_3$ thin film. The work function = $\Phi_{4\text{-CNpyBiI3}}$ was determined from the bias independent secondary electron cutoff ($\Phi_{4\text{-CNpyBiI3}}$ = hυ − secondary cutoff

energy = 21.22-16.4 = 4.82 eV) (A). The injection barrier ($\Phi_B$) was determined to be 0.3 eV from the Fermi level (B). The ionization energy (IE) = $\Phi_{4\text{-}CNpyBiI3}$ + $\Phi_B$ = 5.12 eV. $E_F$ : Fermi level, valence band maximum ($VB_{max}$), Conduction band minimum ($CB_{min}$), and electron affinity (EA).

Table 1. Resulting parameters from UPS and UV-vis measurements.

| Semiconductor | Optical Band gap/eV | Work function ($\Phi_S$)/eV | Hole Injection ($\Phi_B$)/eV | SE cut-off/eV | Conduction band to $E_f$ | EI |
|---|---|---|---|---|---|---|
| 4-AmpyBiI$_3$ | 2.1 | 4.92 | 0.8 | 16.3 | 1.3 | 5.72 |
| 4-MepyBiI$_3$ | 2.2 | 5.02 | 0.2 | 16.2 | 2 | 5.22 |
| 4-DmetpyBiI$_3$ | 2.3 | 3.52 | 1.5 | 17.85 | 0.8 | 5.02 |
| 4-CNpyBiI$_3$ | 1.99 | 4.82 | 0.3 | 16.4 | 1.69 | 5.12 |

The memristor devices comprised of thin layers (featuring an electron-withdrawing cation in its crystalline structure) on ITO glasses (which serve as the counter electrode, assumed as a metallic electrode in calculations) were investigated with various metallic electrodes (such as copper (Cu; work function = 4.7), gold (Au; work function = 5.2), aluminum (Al; work function = 4.3), and silver (Ag; work function = 4.26-4.73)). The objective was to determine the most suitable metallic electrode that exhibits stable and pronounced hysteresis loops in the V curves (Figure S3).[33] For a device to function as a memristor with resistant switching behavior, it should have a Schottky-type contact. According to this model, for the p-type semiconductor the work function of the metal should be lower than that of the semiconductor ($\Phi_m < \Phi_s$), but too large a difference may result in too high resistivity of the junction and unstable operation of the device.[34] Crystalline structures with distinct bismuth iodide fragments are described in the table (Table 1), and as indicated by the UPS results, the resulting complexes are p-type (except 4-DmapyBiI$_3$, which according to UPS measurements, is an n-type semiconductor), probably stemming from an excess of pyridinium iodide species. This conclusion is substantiated by UPS measurements, clearly indicating that for 4-DmapyBiI$_3$ the Fermi level in collated just below the lower edge of the conduction band, whereas for other compounds is in located 0.2 – 0.8 eV above the upper edge of the valence band (Figure 4).

It is crucial to note that Ag electrode exhibited prominent pinched-shaped hysteresis loops; however, it proved to be unstable, undergoing rapid degradation due to reaction with iodide-based semiconductors, and disappearing from the device surface within one day. In place, there were blank areas with a yellowish film (Figure S3), possibly indicating the formation of AgI. On the other hand, the Cu electrode demonstrated greater stability under ambient conditions. Even after one month, the results remained reproducible, showcasing consistent clockwise-counterclockwise I-V curves. In this case a good device-to-device reproducibility was also observed (Figure S4).

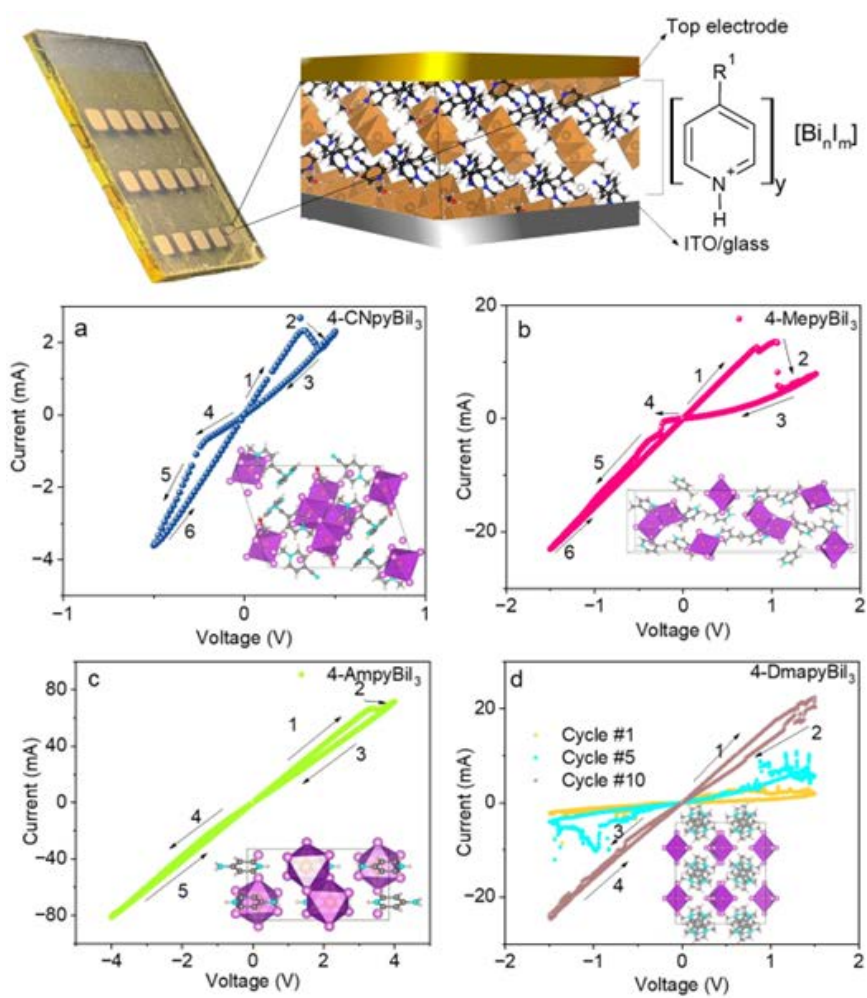

**Figure 5.** (a-d) Current-voltage (IV) characteristics of Bismuth complexes, unit cell, are represented as inset. (d) different cycles (no. 1, 5, and 10) are represented for 4-DmapyBiI$_3$. At the top scheme R$^1$ = -CN, -CH$_3$, -NH$_2$, and –(CH$_3$)$_2$N-).

The complexes exhibited hysteresis loops in different voltage ranges (Figure 5). Specifically, 4-CNpyBiI$_3$ displayed an intense loop at low voltages ranging from +0.5 to -0.5 V. In the case of 4-MetpyBiI$_3$, the loops appeared at ±1.5 V, while for 4-AmpyBiI$_3$, voltages exceeding ±3 V were necessary to observe the loops. This voltage trend aligns with the electron-donating nature of functional groups in the pyridine moieties, where the order of effectiveness is 4-CNpyBiI$_3$ < 4-MepyBiI$_3$ < 4-AmpyBiI$_3$ (Figure 5). In particular, the I-V curve of 4-DmetpyBiI$_3$ proved to be unstable at a scan rate of 500 mV/s, exhibiting irregular changes in each cycle and undesired resistance fluctuations. Consequently, this sample was not further investigated. The shape of hysteresis loops is consistent with the p-type character of studied semiconductors.

On the basis of these findings, copper was selected as the practical electrode for the subsequent phases of the study. Issues arose with devices with Au and Al electrodes, as they did not show observable pinched-shaped loops, instead as ohmic contacts without resistive changes (Figure S3). In the case of 4-CNpyBiI$_3$, the distinctive pinched shape was noticeable within a low voltage range of ±0.5 V, while for 4-MepyBiI$_3$, it appeared at ±1 V. For 4-AmpyBiI$_3$, an increase in voltage to ±4 was required to discern the characteristics of memristive devices (Figure 5). This behavior is consistent the resistive switching mechanism based on Schottky barrier height modulation.[35, 36]

The significance of the organic cation in the perovskite-like structure has previously been underestimated,[37] but our observations underscore that gradual alterations in functional groups on the organic cation, transitioning from electron-donor to electron-acceptor groups, can influence the performance of resulting devices. Even small changes in the cation structure results in modification of hydrogen bond network in the lattice, the local geometry of bismuth centres, and long distance interaction, which in consequence translates into differences in band gap energies, work function, charge carrier mobilities and other electronic parameters. This in turn affects the structure and properties of metal-semiconductor interface, Schottky barrier height and translates into the performance and stability of memristive devices based thereof. Consequently, this affects the resistive switching process and the shape of the IV plot (switching potential, ON/OFF ratio, symmetry of the I-V curve, etc.). For instance, the bismuth complex containing a functionalised pyridinium cation with a dimethyl amino group as an electron-donating entity proves to be the most unstable device with a low likelihood of reproducing measurements. On the other hand, the 4-cynaopyridinium cation (electron-withdrawing group) yields the most promising results. The IV curve under different scan rates and voltage ranges (Figure S5), the ON-OFF ratio (Figure 6), and the retention of the 4-CNpyBiI$_3$ device were evaluated and compared with other derivatives (Figure S6). The effect of surface area on ON/OFF ratio was studied for two types of electrode (with 1mm$^2$ and 9mm$^2$), the results show increasing of conductivity by increasing the surface area of the electrode (Table S2). This results along with rectifying factor

(Table S3) and asymmetric shapes of IV loops, bring us to this conclusion which interfacial surface id current driving factor in these kind of materials and devices.

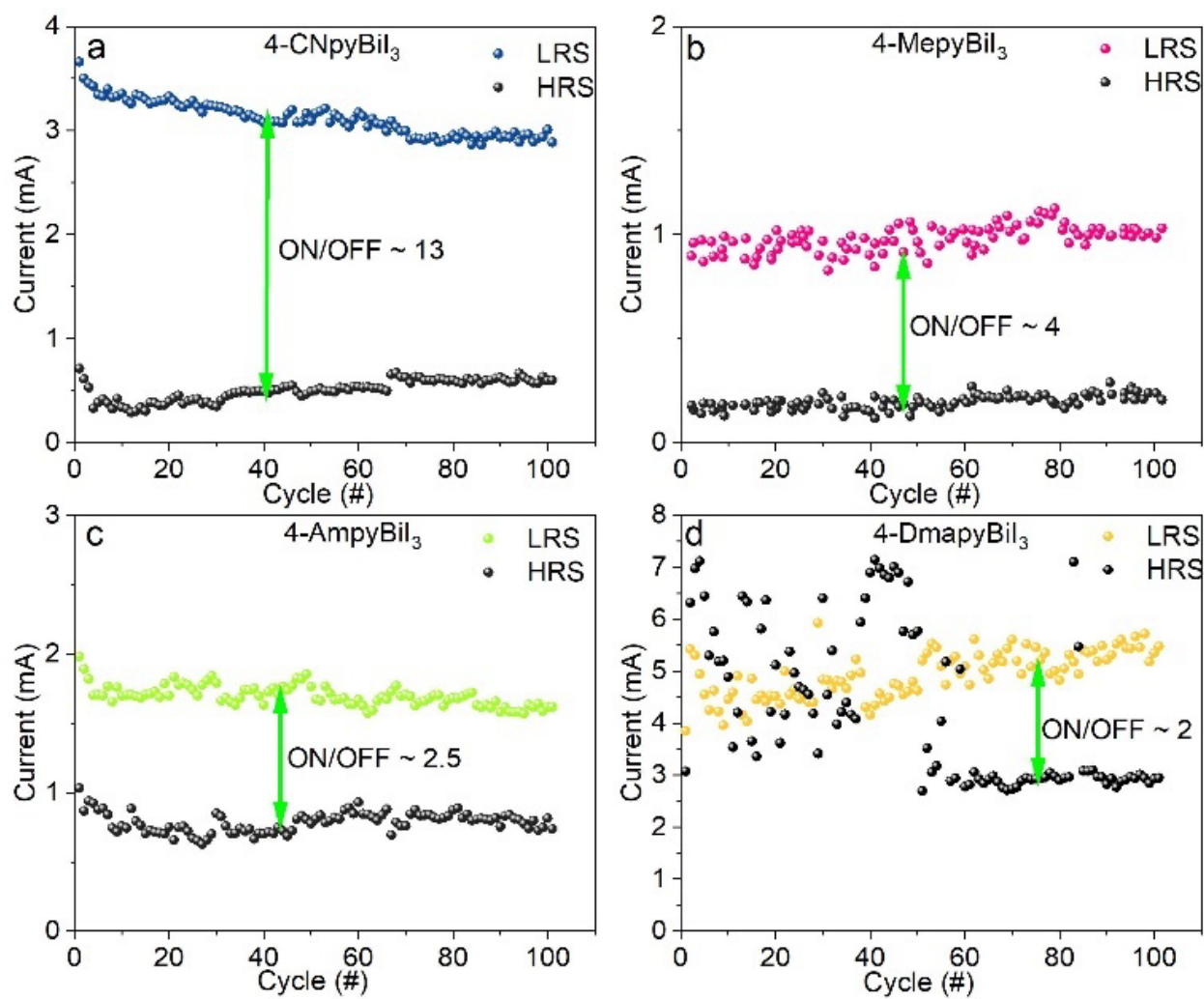

**Figure 6**. On-off ratios of the devices made of a) 4-CNpyBiI$_3$ and b) 4-MepyBiI$_3$, the range of ±2 V was chosen as the set and rest voltages for a and b and (±4 V for c) and the read point was + 0.2 mV, d) unstable on-off ratio in 4-DmapyBiI3 with ±2 V as the set and rest values. The width of all pulses was 0.1 s.

The characteristics observed in the current-voltage (IV) curves provide crucial insights into the resistance switching mechanisms of the device, which consists of pyridinium-based bismuth complexes sandwiched between two distinct electrodes, one on top and the other at the bottom. If the dominant mechanism were filamentary-type resistive switching, the IV curve would exhibit a sudden increase in set and reset points, accompanied by a symmetric hysteresis pattern. However, in this study, the IV curves exhibit an asymmetric appearance, indicating the prevalence of interface-type interactions between the sandwiched materials and each electrode positioned on top and bottom.[35, 36]

Under positive voltages, the transition takes place from LRS to HRS, termed the RESET process. This phenomenon is plausibly explained by the application of positive voltages that induce ionic movements and the redistribution of ions within solid structures combined with modulation of the Schottky barrier heights (at least for 4-MepyBiI$_3$, for which an asymmetric hysteresis loop has been observed, (Figure 5b).[38] Consequently, a resistance switch from HRS to LRS is observed during the SET process when negative voltages are applied to the working electrode. Current-voltage (IV) plots were analyzed over a temperature range of 243 K to 393 K, and the impacts on SET and RESET points are illustrated in Figure 7. The significant variations in the RESET point (LRS to HRS) with changing temperatures suggest potential suppression or activation of ion migration, leading to alterations in conductivity (either reduced or increased).

In devices exhibiting resistance switching, heterogeneous features such as local variations, defects, or impurities in the material composition can give rise to distinct activation energies in different regions of the device. Completion and characterization of these variations in activation energy within a resistance switching device present a challenging task. These variations may originate during the fabrication process or emerge due to ageing of the material and environmental influences. A comprehensive review of the existing literature suggests that at low voltages, the metal-semiconductor interface predominantly plays a role. However, at higher voltages, the bulk body can influence the conduction path.

The slopes observed in log-log plots of current (I) versus voltage (V) in resistive switching devices serve as a means to identify the dominant charge-transport mechanisms, aiding researchers in understanding and characterizing device behavior. Various mechanisms exhibit distinctive slopes, including Schottky barrier (SB), space-charge limited current (SCLC), and trap-assisted tunneling (TAT).[39]

TAT involves carriers being captured by trap energy states within a material's bandgap and subsequently tunneling through these traps. This phenomenon significantly influences the electrical conduction and device behaviour, particularly at low temperature (e.g. -30 °C), especially in crystals with 0D ionic fragments in their morphology (Figure S7),[29] as observed in the reset process of crystals such as 4-CNpyBiI$_3$, 4-MetpyBiI$_3$, and 4-DmetpyBiI$_3$ (Figure 7).

Figure 7 illustrates the changes in a double logarithmic plot at various temperatures during the RESET process, which occurs within the positive voltage range of 0 V → +2 V and +2 V → 0 V, as well as the SET process under negative bias, ranging from 0 V → -2 V and -2 V → 0 V. Ohmic and TAT are identified as the primary resistance-determining factors, both associated with electronic conductions. Specifically, the 1D morphology of ionic fragments in 4-AmpyBiI$_3$, characterized by a low void percentage and a well-ordered crystal structure in its packed unit cell,[29] exhibits Ohmic conduction as the predominant mechanism at low temperatures (Figure S8). However, as the temperature increases to room temperature or higher, TAT becomes observable, accompanied by the emergence of resistive switching regions.

In the case of 4-CNpyBiI$_3$, 4-DMapyBiI$_3$, and 4-MepyBiI$_3$, which possess 0D ionic fragments of Bi-I bonds in their structure, switching loops with a higher surface area are observed at low temperatures (-30 °C), attributed to ionic conduction[29] (Figure S6). At higher temperatures (150 °C), the I-V plot in the two former compounds shows narrower hysteresis loops, and the log-log plot aligns more closely with an ohmic-like trend. With an increase in the device temperature from room temperature to 150 °C, ionic conductivity is facilitated, which results in narrowing of the hysteresis loop – there is no noticeable latency between the variation of potential and the distribution of ions in the solid film and the only contribution to resistive switching comes from charge trapping at the interface. In the case of 4-AmpyBiI$_3$, the only structure with one-dimensional polymeric chains, the situation is different. At low temperatures hysteresis is negligible due to dominating ohmic conductivity of the material. At higher temperatures, however the increased concentration of lattice defects is possible. Therefore observed hysteresis opening at higher temperatures may be attributed to increased contribution of SCLS process, decreased contribution of ohmic conductivity (due to higher phonon scattering) and maybe also increased ionic mobility.

In a log-log plot, the slope for trap-assisted tunneling may not adhere to a fixed value and can vary significantly. Variation in slope could result from increased ionisation of traps as the temperature increases, facilitating easier tunneling of the carriers through these traps.

Log-log plots were generated under negative bias conditions and are presented in Figure S8. For 4-MepyBiI$_3$, the presence of SCLC (I ∝ V$^2$) was identified. Sequentially, Ohmic-like regimes (I ∝ V) and TAT with slopes higher and lower than 2 were observed. In the case of 4-CNpyBiI$_3$, positive bias revealed three main observable mechanisms: Ohmic (LRS: 0-1V; HRS: 0.97-0), SCLC (~1.49-

0.98V), and TAT (~1-1.22 V). Ohmic voltage ranges dominated in both the HRS and LRS regimes at 298 K. However, in negative bias, only Ohmic (~0 to -0.13 V, -1.5 to 0) and SCLC (-0.13 to -0.3 and -0.31 to -1.5) were prevalent, with SCLC being the dominant feature in the HRS regime. For 4-MepyBiI$_3$, positive voltages exhibited all three transport paths, while in negative bias, Ohmic and TAT (in the SET process) were the prominent mechanisms. Constant–potential conductivity measurements indicate a significant contribution of ionic transport in the total conductivity in these materials. These measurements indicate the activation energies of 72, 76, 80, and 130 meV for 4-CNpyBiI$_3$, 4-AmpyBiI$_3$ 4-MepyBiI$_3$, and 4-DMapyBiI$_3$, respectively).[29]

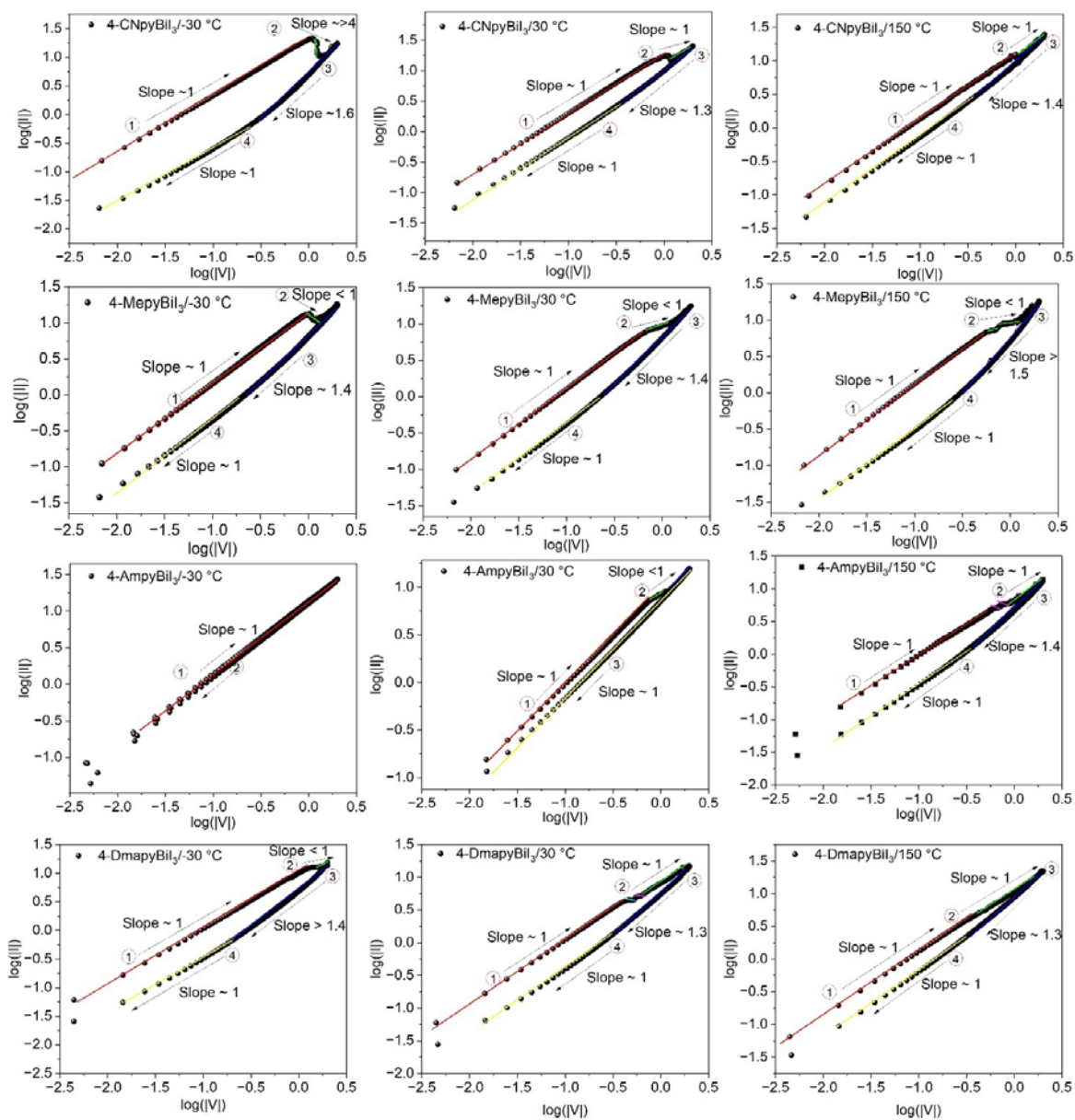

**Figure 7**. Double logarithmic scales of IV values of 4-CNpyBiI$_3$, 4-MetpyBiI$_3$, 4-DmetpyBiI$_3$ and 4-AmpyBiI$_3$ in RESET process.

**Effect of the functional group on plasticity and neuromorphic studies**

Devices based on 4-CNpyBiI$_3$ and 4-MepyBiI$_3$ operate within a narrow voltage window (i.e. undergo a resistive switching at low potentials), and measurements of the on/off ratio reveal two stable resistance states with a relatively low ratio (~ more than 2 and less than 13). However, there are two discernible HRS and low-resistance states LRS, suggesting their potential suitability in artificial neural networks. Investigating long-term potentiation (LTP) and long-term depression (LTD) features, excitatory postsynaptic current (EPSC) and STDP using solid-state neuromorphic devices is an intriguing avenue.

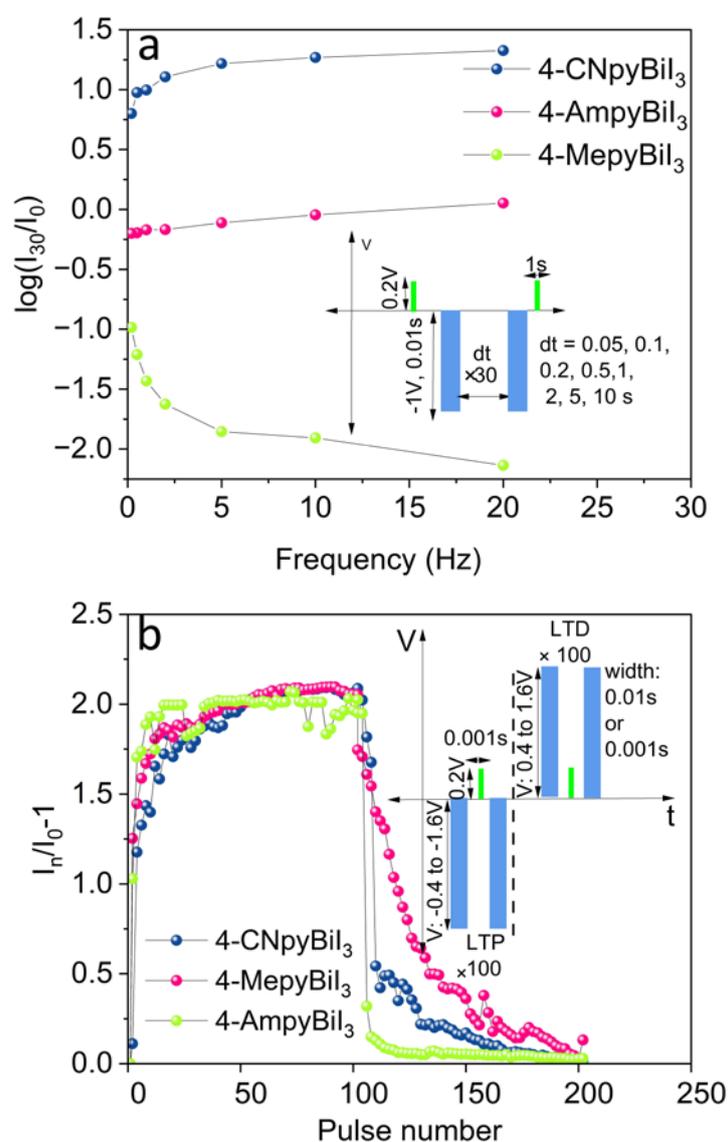

**Figure 8**. a) Frequency-dependent EPSC of 4-CNpyBiI$_3$, 4-MepyBiI$_3$, 4-AmpyBiI$_3$. The inset represents the applied procedure, b) Potentiation-depression, room temperature, 1 ms width of pulses with 1.6 V amplitude.

Figure 8a shows the EPSC response to various-frequency spike trains. Thirty sequential spikes were put on the procedure with different time width (0.05 to 10 s), while the read point was fixed at 0.2 V. The EPSC gain is the difference between the first and last absolute EPSC values ($I_0$ and $I_{30}$). A higher frequency causes the EPSC peak values to rise along with the spike counts. The EPSC gain rises from 6.3 to 21.7 when the frequency increases from 1 to 100 Hz in case of 4-CNpyBiI$_3$, which shows that the device has frequency-dependent synaptic response. In this framework, EPSC investigations show negligible frequency dependence of devices fabricated from 4-AmpyBiI$_3$ and 4-MepyBiI$_3$ materials compared to the former sample.

The investigation involved one-pulse examinations on the fabricated devices, exploring long-term potentiation (LTP) and depression (LTD) at different voltages, with results displayed in Figure 8b, take into account that the applied devices show clockwise IV curves, which means negative voltages cause potentiation and positive pulse cause depression. For each depression or potentiation, a series of consecutive pulses (100 cycles) with varying amplitudes (±0.4 V to ±1.6 V) and different widths (0.01 s and 0.001 s) were selected and the read current at 0.2 V was recorded. In the case of 0D 4-CNpyBiI$_3$ the applied short pulses (0.001 s) result in gradual and almost linear increase of current at applied voltage (-0.8 V) and gradual depression. In case of 4-MepyBiI$_3$ two linear trends in potentiation were observable (at -0.8V and -1.00 V), but just at +1V gradual depression appeared. However in case of 4-AmpyBiI$_3$ the potentiation and depression both were nonlinear in all applied voltages (Figure S9). All these phenomena are related to the different conductivity mechanisms studied in log-log plots (Figure 7, Figure S8).

In the case of LTP plots, an increase in voltages higher than -1V demonstrated higher conductivity changes compared to lower voltages (-0.6, -0.8 and -1). The conductivity ratio between In/In-1 approached 1, indicating that potentiation is independent of applied voltages beyond -1V, and non-linearity increased. Consequently, application of short pulses with 0.001 sand -1.6 V pulses resulted in a sharp increase of conductivity with an increase in the number of pulses, aligning with the synaptic weight increase observed in biological synapses. The rate of conductivity rise is almost the same for all three (4-CNpyBiI$_3$, 4-MepyBiI$_3$, 4-AmpyBiI$_3$) studied materials (Figure 8b), with 4-CNpyBiI3 showing slightly slower response.

All these time responses can be fitted with pseudoexponential functions for both potentiation (biexponential, 1) and depression (monoexponential, 2):

$$W_{potentiation} = A_1 e^{-\frac{n}{\tau_1}} + A_2 e^{-\frac{n}{\tau_2}} + W_0$$

$$W_{depression} = Ae^{-\frac{n}{\tau}} + W_0$$

where n is the pulse number, τ, τ1, and τ2 are pseudo-time constants and W0 is the asymptotic synaptic weight value. Pseudo-time constants are collated in Table 2.

Table 2. Pseudo-time constants fitted for potentiation ($\tau_1$, $\tau_2$) and depression ($\tau$).

| Compound | $\tau_1$ / ms | $\tau_2$ / ms | $\tau$ / ms |
|---|---|---|---|
| 4-AmpyBiI$_3$ | 1.14±0.1 | 12±6 | 6.1±0.7 |
| 4-MepyBiI$_3$ | 0.6±0.05 | 23±2 | 25.6±0.8 |
| 4-CNpyBiI$_3$ | 2.7±0.4 | 33±8 | 12±1 |

Taking into account the character of the substituent it can be concluded, that materials containing cations with electron- donating functionality are fast responding, whereas the electron-withdrawing functionalities in the structure result in slower response. It is intuitively correct: in the case of metal – p-type semiconductor an increase of conductivity (i.e. a decrease of the Schottky barrier height) required trapping of electrons at the interface. Electron withdrawing groups may act as competitors, resulting in is electron trapping not at metal-induced gap states, which is necessary for reduction of the Schottky barrier height, but in the defects in the bulk of the material. Electron donors (especially the amino derivative), on the other hand, do not participate in bulk electron trapping, which in consequence results in faster response.

These results are consistent with the asymmetric shape of the I-V curve. The current findings indicate that the artificial synapse effectively mimics the synaptic increment in line with the corresponding biological analogues. Consequently, the results have led us to apply voltages equal or higher than ±1 in pulses for spike timing-dependent plasticity experiments (STDP), resulting in recognizable conductive changes.

To examine the dependency of plasticity on time and waveform changes on synaptic weights in devices composed of 4-CNpyBiI$_3$ and 4-MepyBiI$_3$, we selected pre- and post-pulses consisting of the triangle, sawtooth and square waveforms (Figure 9). The STDP was implemented by applying pre- ($V_{pre}$) and post-pulses ($V_{post}$) at different time intervals, where the time difference was sequentially positive (Δt = $t_{post}$ - $t_{pre}$ > 0). The procedure to start and resetting the device is illustrated as inset, while changes in synaptic weights are depicted in Figure 9.

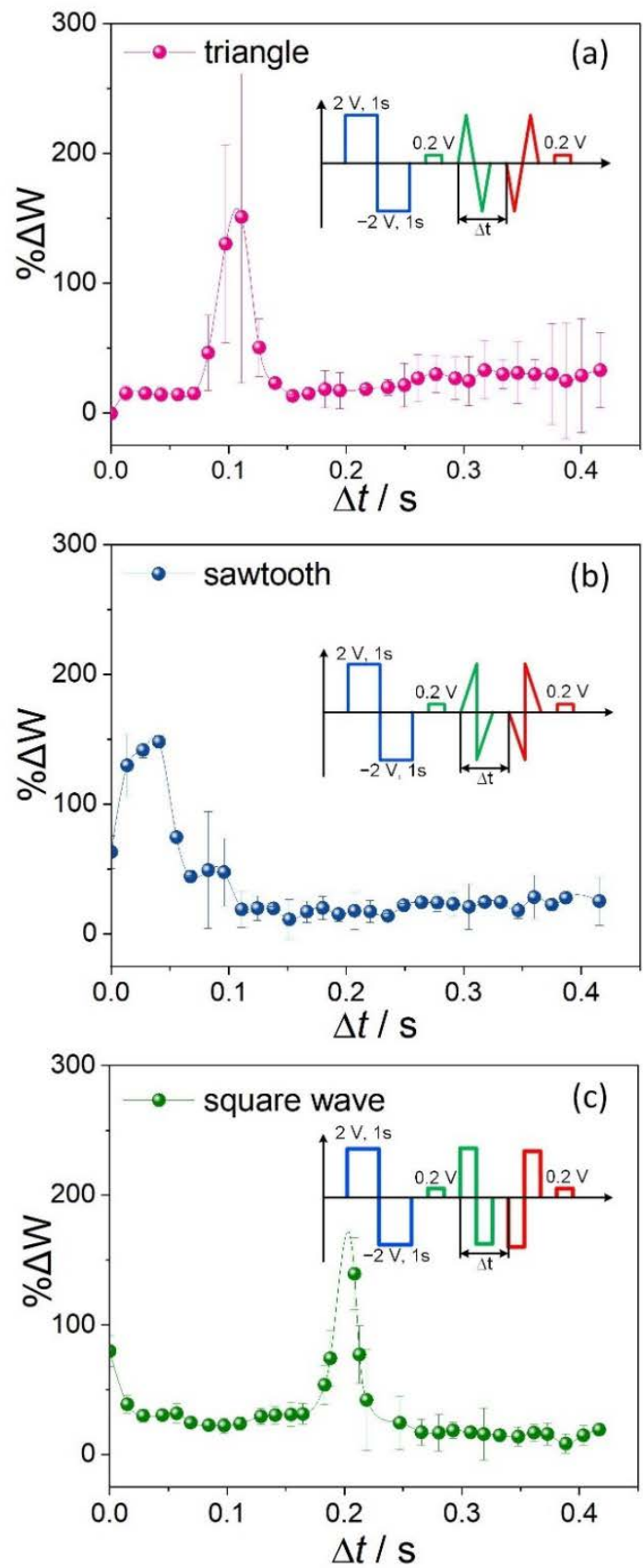

Figure 9. a) triangle-type, b) sawtooth pulse and c) square pulses and their corresponding synaptic weight changes in 4-CNpyBiI$_3$.

Unfortunately, the procedure employed in most of the papers in this field lacks clarity, particularly regarding the impact of different-shaped pulses with positive or negative polarities, and the influence of waveform shape and interval between waveforms on the Hebbian learning rule has not been adequately elucidated. Consequently, the final results demonstrate an asymmetric Hebbian learning rule.[14, 40] However, the results in this paper unequivocally highlight how changes in waveform over time can significantly affect conductivity and plasticity.

Two synchronized arbitrary waveform generators (GEN) control two voltage-controlled source-measure units, configured as voltage-controlled voltage sources with current monitoring circuit (SMUs) provide voltage input and current measurement through the device under test (DUT). Both GEN and SMUs are internal modules of the Keithey SCS-4200 system (Figure 10). Waveforms and timing of GEN, as well as current reading at SMUs are performed owing to internal computer-controlled interface and analog-to-digital converter (not shown on figure).

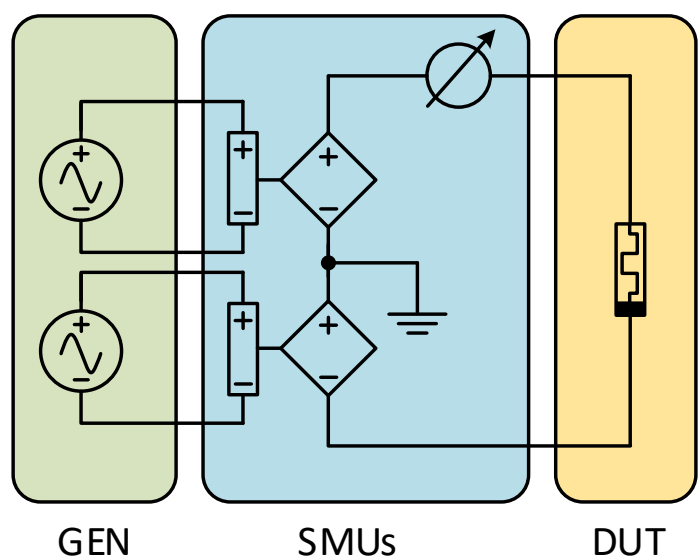

Figure 10. The simplified block diagram of the circuit for plasticity measurements.

Figure 11a illustrates four distinct sequences of sawtooth pulses, each with different setting polarities (referred to as #1, #2, #3, and #4). These pulses were applied and the results are depicted in Figure 11b-e, where $V_{pre}$ = ±1 and $V_{post}$ = ±1 serve as the two wings of STDP behavior in neuromorphic devices. The superimposition of pulses in various regions produces the final pulse shape in memristor devices, leading to constructive or destructive effects that induce changes in conductivity or synaptic weights through pulse techniques (Figure S10). When Δt equals or exceeds the duration of each applied pulse, which means there is no overlap between pre- and post-pulses, the synaptic weight changes are approximately zero (Figure S10). Based on these preliminary findings, these insights were applied in two-pulse experiments. In these experiments, the ITO electrode was considered to be the equivalent of the postsynaptic neuron, and the Cu electrode was considered the presynaptic one. The shapes, amplitudes, and time intervals of the pulses are critical factors influencing the changes in conductance ($\Delta G = G - G_0$), where $G_0$ and $G$ represent the conductance before and after simulated synaptic events with different time intervals of applied pulses in the device, respectively. Results in this work prove that shapes of conductance and synaptic weights are depended on the shape and polarity of applied spikes. Following, we represent the results of two sawtooth pulses (upward pulse : downward pulse with amplitude +1V: -1V; and -1V: +1V) with the same amplitude but mirror shapes in 4-CNpyBiI$_3$ as applied materials in the memristor device, therefore, the resulting STDP has mirror shape (Figure 11).

Figure S10 display three primary regions with respect to the time difference of pulses ($\Delta t$), influencing the ultimate voltage shape ($V_{pre} - V_{post}$), a crucial factor in determining the final synaptic weights. As $\Delta t$ approaches zero, the conductance of the device reaches its maximum ($G_{max}$) due to the higher voltages resulting from the superposition of spikes. However, a time difference exceeding the pulse lifetime (Figure S10) shows the device's intact conductance without interference between pulses. In this context, the changes in synaptic weight changes ($\Delta W$), correlated with the conductance changes ($\Delta G$) in artificial neuromorphic devices ($\Delta W = (G - G_0)/G_0$), exhibit a symmetric Hebbian plot, particularly potentiated around $|\Delta t| \sim 0$. By increasing the pulse amplitude to (±2 V), negligible changes in synaptic weights are observable (Figure S11).

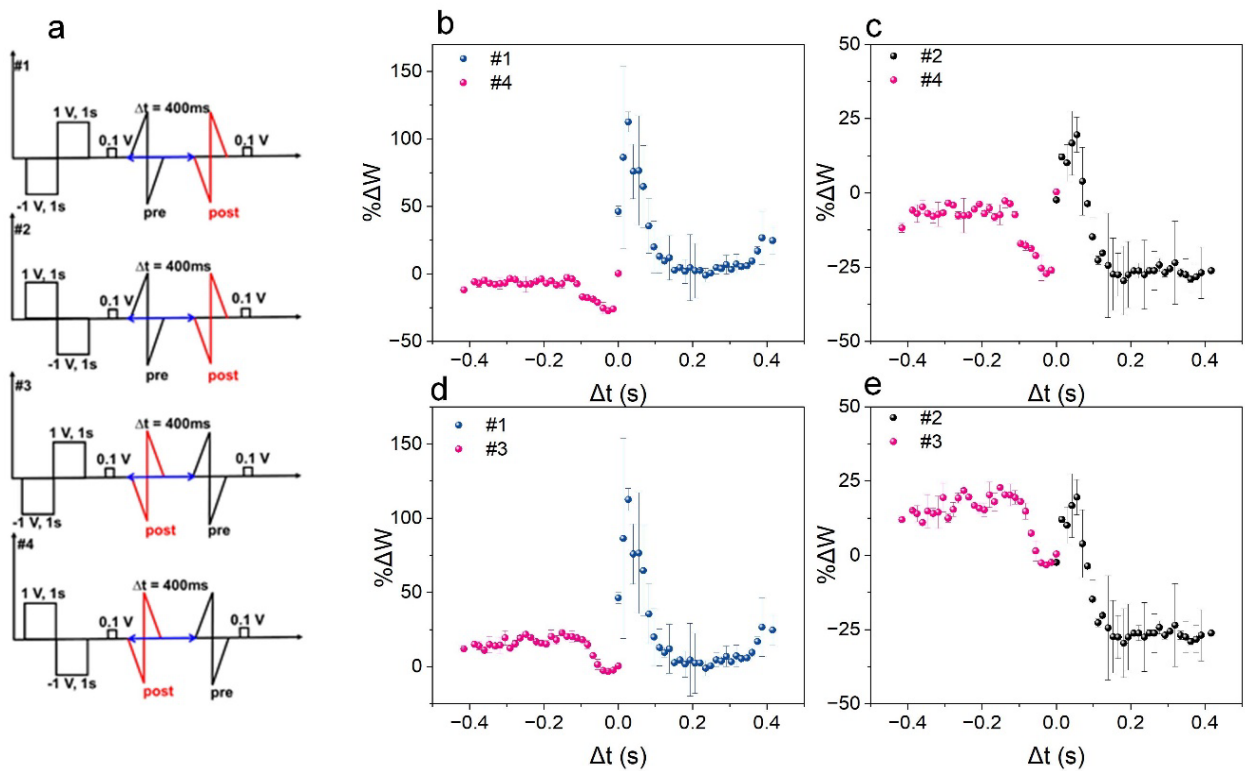

**Figure 11.** (a) Four types of sawtooth pulse sequences as #1 #2 , #3 and #4 are depicted. STDP of 4-CNpyBiI$_3$ in different pulse polarities b) #1: #4 c) #2: #4, d) #1: #3, e) #2: #3

STDP results for devices made of 4-MetpyBiI$_3$ represent similar trends (Figure S12). The only exception occurred when there was an increase in the amplitude of the pulses in #1, resulting in irregular changes in synaptic weights; however, the depression part (#4) was reproduced successfully (Figure S13).

For 4-AmpyBiI$_3$, as previously discussed, hysteresis loops are not detectable at room temperature, and the weight changes are minimal and highly noisy, making them challenging to represent accurately.

**Experimental**

**Materials and characterisation techniques.**

Absorption spectra of thin layers on ITO glasses were measured in a studyby a LAMBDA 750 UV/vis/NIR spectrophotometer (PerkinElmer Inc., USA). Cross-sectional SEM was measured by FEI Versa 3D double-beam high-resolution scanning electron microscope with accelerating voltage range from 500 V to 30 kV. The thickness of the layers was estimated by DektakXT stylus profilometer (Bruker DektakXT). A LeicaEM ACE600 high-vacuum sputtering machine was employed to sputter thin film metals from two sources of dissimilar metals. Ultraviolet photoelectron spectroscopy (UPS) measurements were conducted using a PHI VersaProbeII apparatus (ULVAC-PHI, Chigasaki, Japan) with a He I line (21.22 eV) from a UHV gas discharge lamp. A -5 V acceleration potential was applied to the sample, resulting in a significantly more pronounced secondary electron cutoff (SE cutoff). The Polos SPIN150i spin coater was applied for the thin layer fabrication process, the optimized procedure for preparation of thin layer preparation.

I-V plot measurements, on-off ratio, memristor state retention, and various plasticity measurements were conducted using an SP-150 potentiostat (BioLogic, France) in a two-electrode configuration. The working electrode (WE) was connected to the top metal electrodes on thin layer, while counter electrode (CE) and reference electrode (RE) were grounded and connected to the ITO layer. spike timing-dependent plasticity (STDP) were investigated using the SCS-4200 system (Keithley, USA) with two SMUs connected to a metal electrode and ITO substrate, respectively. Spectroscopic ellipsometer SENresearch 4.0 - model 850 (SENTECH Instruments GmbH, Germany) was applied to evaluate both thickness and optical properties of semiconducting layers.

**Preparation methods**

Synthesis of bismuth iodide-derived semiconductors has been reported an a previous paper.[29] CIF files are attached as Supplementary Data and the elemental analysis for each structures is as follows: 4-CNpyBiI$_3$ (C$_{14}$H$_{16}$Bi$_2$I$_8$N$_4$O): yield 90%, elemental analysis found as C, 6.84%; H, 0.69%; N, 2.3%; 4-MepyBiI$_3$ (C$_{18}$H$_{24}$Bi$_2$I$_9$N$_3$): yield 98%, elemental analysis found as C, 8.85%; H, 0.964%; N, 1.6%; 4-AmpyBiI$_3$ (C$_5$H$_7$BiI$_4$N$_2$): yield 95%, elemental analysis found as C, 9.14%; H, 0.97%; N, 3.73%; 4-DmapyBiI$_3$ (C$_{28}$H$_{44}$Bi$_2$I$_{10}$N$_8$): yield 92%, elemental analysis found as C, 14.49%; H, 1.89%; N, 4.68%.

The device preparation was proceeded by preparation of thin layers from an optimised concentration (300 mg of pyridinium-bismuth salts in 1 mL dimethylformamide (DMF). Spin coating procedures consist of two sequential steps with spin acceleration 1000 rpm: step 1 with spin speed 1000 rpm for 10 s and step 2 with spin speed 2000 rpm for 31 s which resulted in uniform layers.

**Conclusions**

The devices with nanolayers of bismuth iodide complexes sandwiched between Cu and ITO electrodes have been fabricated. UPS results show these crystalline materials, with varying cationic compositions and ionic fragment dimensionalities, are p-type semiconductors with band gaps under 2.3 eV. Despite structural and electronic similarities voltages required to observe resistive switching in 4-CNpyBiI$_3$, 4-MepyBiI$_3$, and 4-AmpyBiI$_3$ varied significantly, reflecting the influence of functional groups within cationic moieties on switching processes. Using a double logarithmic plot at different temperatures, Ohmic and TAT conductivities has been identified as key electronic conduction mechanisms. At low temperatures, 4-AmBiI3 shows ohmic conduction due to its 1D ionic fragments, shifting to TAT and resistive switching as temperature rises. In contrast, 4-CNpyBiI$_3$, 4-DMapyBiI$_3$, and 4-MeBiI$_3$ (with 0D ionic fragments) exhibit ionic conduction switching loops at low temperatures and ohmic-like behavior at 150°C. Higher temperatures facilitate carrier tunneling by increasing trap ionization. The effect of functional groups (electron donors vs. electron acceptors) on plasticity has been analyzed, indicating significant influence of substituents on long-term depression and potentiation with varying pulse amplitudes and widths. In 0D 4-CNpyBiI$_3$, short pulses (0.001 s) caused a gradual current increase at -0.8 V and gradual depression. For 4-MepyBiI$_3$, linear potentiation trends appeared at -0.8 V and -1.00 V, while +1V caused gradual depression. In 4-AmpyBiI3, potentiation and depression were nonlinear across all voltages. To simulate pre- and postsynaptic spikes in neural network and Hebbian plasticity, among three different applied waveforms right-angle pulse represent promising results in STDP experiments. The height and time of pulses were adjusted (dt = 12, V = ± 1 or ±2). The results show that the amplitude of the applied pulses is in close contact with the structure of the applied materials. Pulses with small height in the case of 4-CNpyBiI$_3$, 4-MepyBiI$_3$ showed high synaptic weight percentages, however in case of 4-AmpyBiI$_3$, the changes were low and negligible.

This study shows the role of substitution effects in cationic sublattice in a series of iodobismuthates, which affects the stability of the devices (indicating that strong electron donating groups, like –N(CH$_3$)$_2$, lead to instability) as well as ON/OFF ratio, amplitude of synaptic weigh changes in neuromorphic experiments and dynamics of switching. These materials provide also much better stability and lower toxicity as compared with lead halide perovskite devices.

## Author contributions

G Gisya Abdi: Conceptualization, Data curation, Formal Analysis, Investigation, Methodology, Resources, Writing – original draft, , Tomasz Mazur: Software, Investigation, Methodology; Ewelina Kowalewska: Investigation, Methodology; Andrzej Sławek: Investigation, Software development; Mateusz Marzec: Investigation, Formal Analysis; Konrad Szaciłowski: Funding acquisition, Conceptualization , Investigation, Writing – review & editing, Project administration, Supervision.

## Conflicts of interest

There are no conflicts to declare.

## Data availability

The data supporting this article have been included as part of the Supplementary Information.

## Acknowledgements

The authors acknowledge the financial support from the Polish National Science Centre within the OPUS programme (grant agreement No. UMO-2020/37/B/ST5/00663) and AGH University of Science and Technology within the program "Excellence Initiative-Research University".

## Notes and references

# Supporting Information

**Table S1.** Extracted parameters from the fitted experimental data to the introduced model for the fabricated devices from pyridinium-based bismuthates samples as Tauc-Lorentz layer.

| Name of the absorber layer | Thickness of Roughness | Thickness of absorber layer | n | k |
|---|---|---|---|---|
| 4-AmpyBiI3 | 9.79 nm | 202.78 nm | 2.13 | 0.0204 |
| 4-MepyBiI3 | 13.98 nm | 150.79 nm | 2.33 | 0.0468 |
| 4-DmapyBiI3 | 20.49 nm | 164.06.45 nm | 2.1170 | 0.00068 |
| 4-CNpyBiI3 | - | 169.16 | 1.5648 ($n_{xy}$ =1.6358; $n_z$ = 1.4226) | 0.38812 ($k_{xy}$ =0.3363; $k_z$ = 0.4917) |

**Table S2.** Effect of the surface area of the electrode on LRS (ON) and HRS (OFF) states.[a]

| Entry | Surface area/1 mm$^2$ | | Surface area/9 mm$^2$ | | Ratio ON[b]/ON[c] | Ratio OFF[b]/OFF[c] |
|---|---|---|---|---|---|---|
| | Current (mA)/ON | Current (mA)/OFF | Current (mA)/ON | Current (mA)/OFF | | |
| 4-CNpyBiI3 | 2.58±0.1 | 1.46±0.1 | 3.00±0.1 | 2.03±0.2 | 1.16 | 1.39 |
| 4-MepyBiI3 | 1.92±0.1 | 1.42±0.07 | 3.08±0.2 | 1.84±0.1 | 1.6 | 1.29 |
| 4-AmpyBiI3 | 2.44±0.1 | 1.94±0.3 | 5.53±0.03 | 3.24±0.4 | 2.27 | 1.64 |

[a]The average amount of 5 different measured devices.

[b]surface area 1mm$^2$; [c]surface area 9mm$^2$

**Table S3.** Rectification factor for devices with different surface area of electrode.[a]

| Entry | Surface area/1 mm$^2$ | | Ratio | Surface area/9 mm$^2$ | | Ratio |
|---|---|---|---|---|---|---|
| | Current (mA) at +2V | Current (mA) at -2V | | Current (mA) at +2V | Current (mA) at -2V | |
| 4-CNpyBiI3 | 30.74±1 | 39.9±0.5 | 1.3 | 24.47±1 | 33.03±0.5 | 1.35 |
| 4-MepyBiI3 | 38.34±0.8 | 48.02±1 | 1.25 | 31.47±0.5 | 39.05±1 | 1.24 |
| 4-AmpyBiI3 | 78.65±1 [b] | 104.24±1.5 [c] | 1.33 | 75.48±1.5 [b] | 83.92±1 [c] | 1.11 |

[a]The average amount of 5 different measured devices.

[b]measured at +4V; [c]measured at -4V

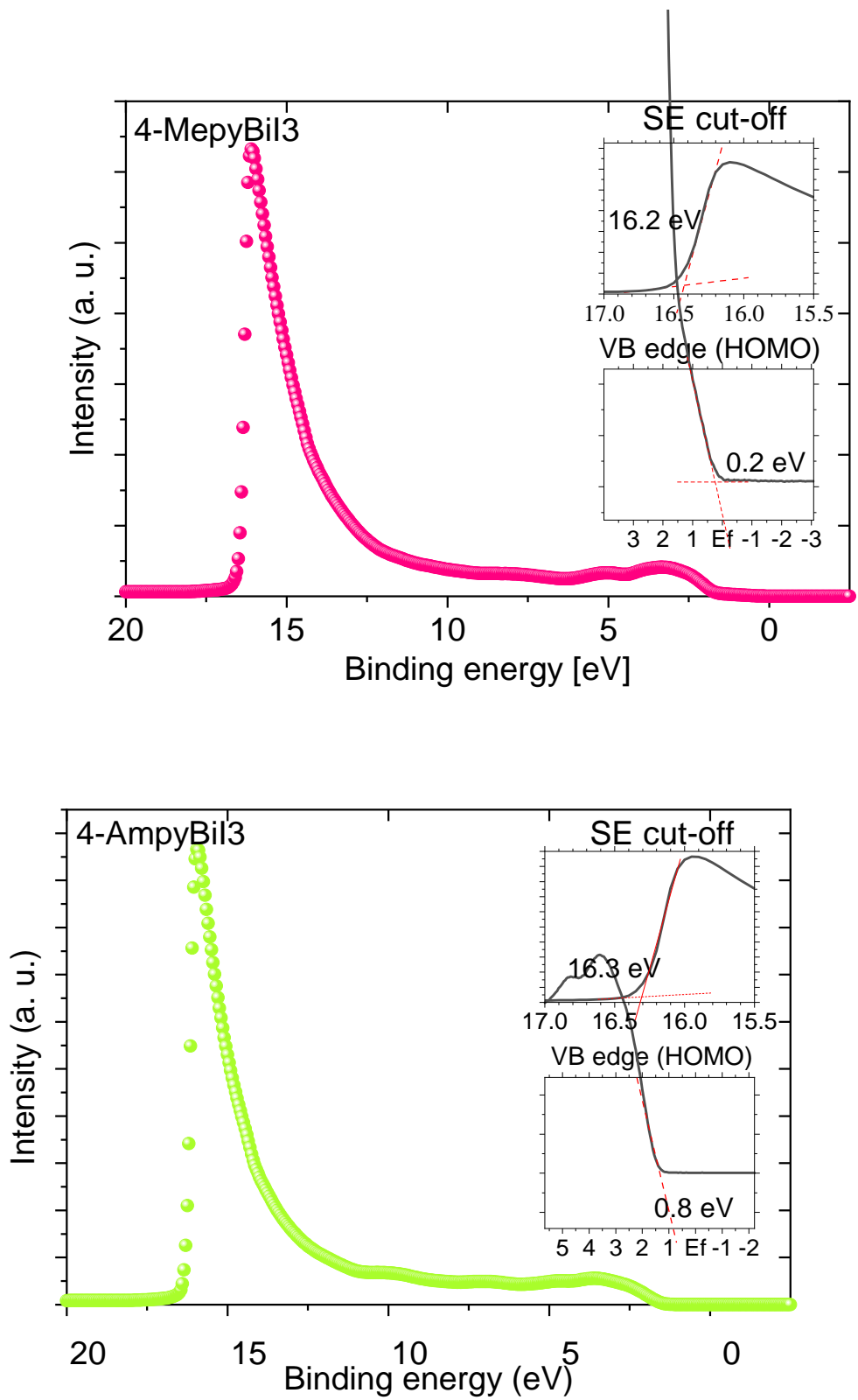

**Figure S1.** UPS spectra of thin films on ITO glasses a) 4-MetpyBiI$_3$ b) 4-AmpyBiI$_3$

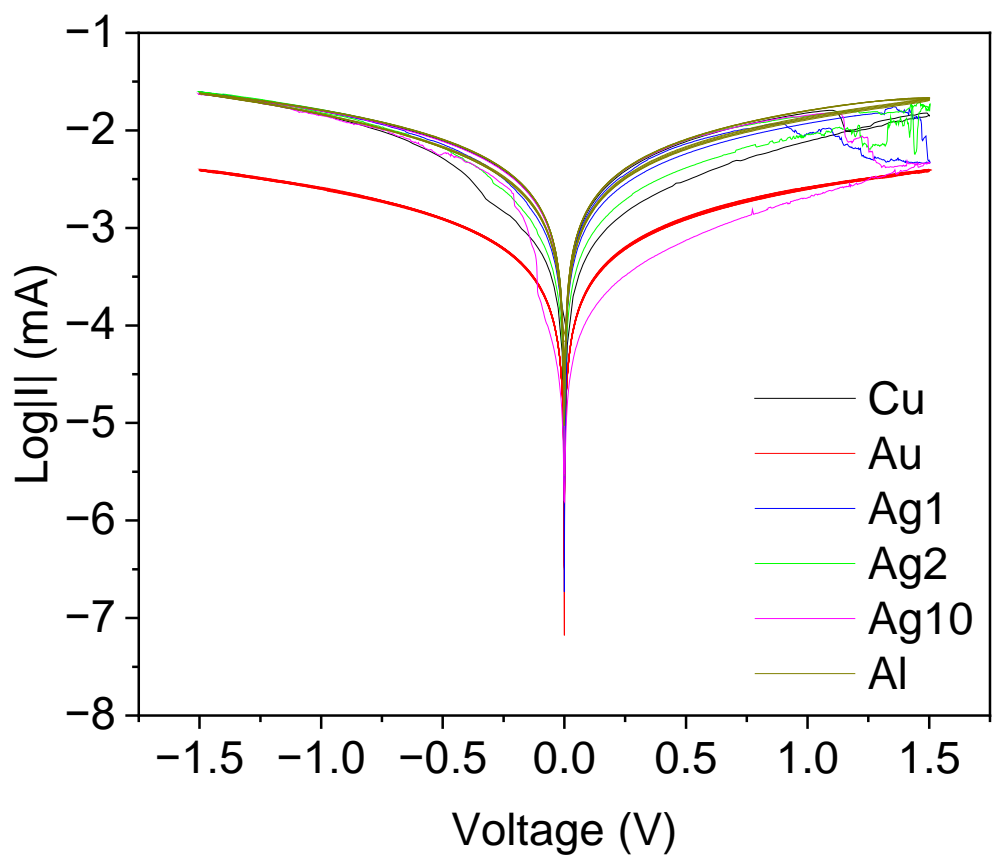

**Figure S2.** 4-CNpyBiI$_3$/ITO glasses with different metal electrodes as top electrode.

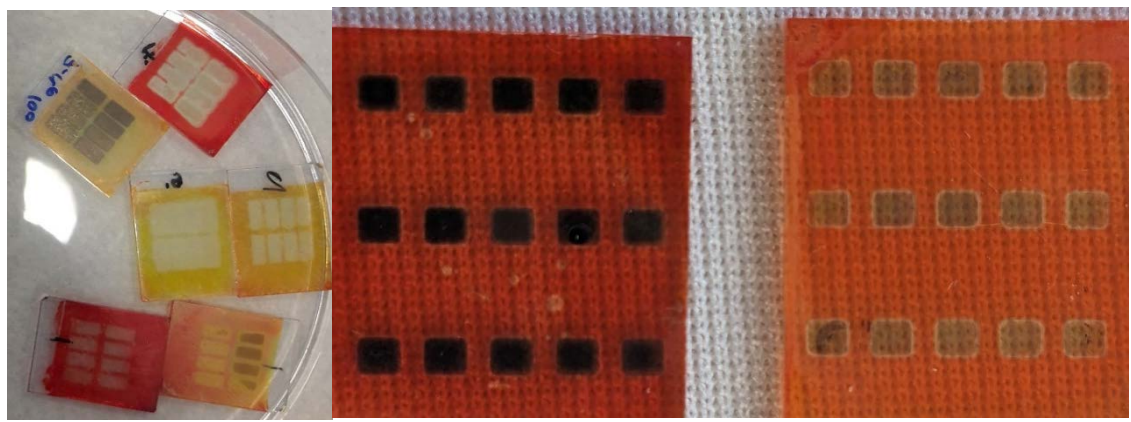

**Figure S3.** Disappearing Ag electrodes after one day from the surface of thin layers of the Bi-complexes on ITO/glasses.

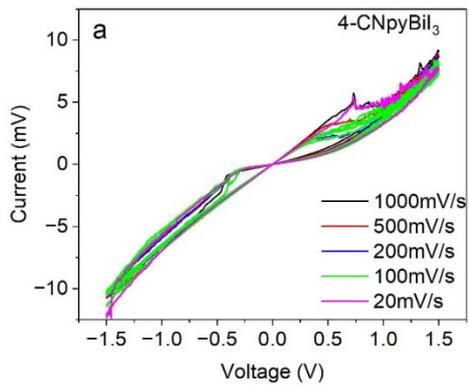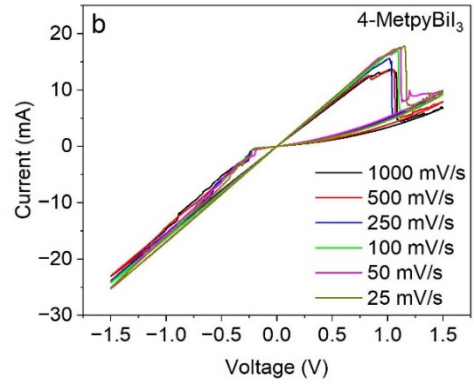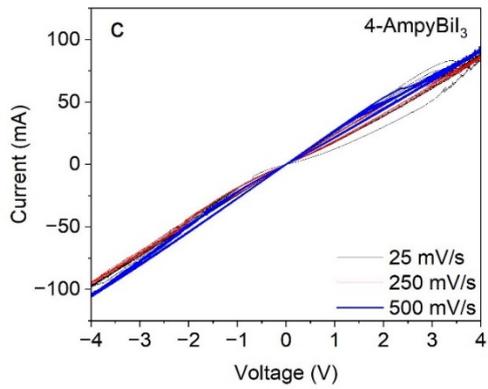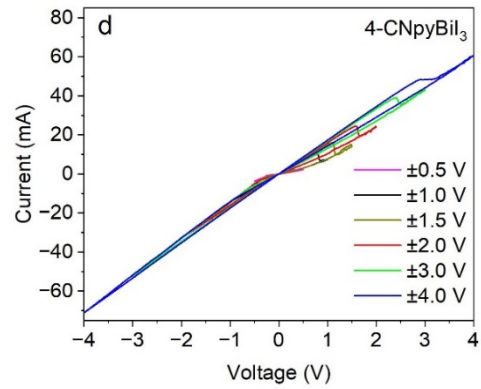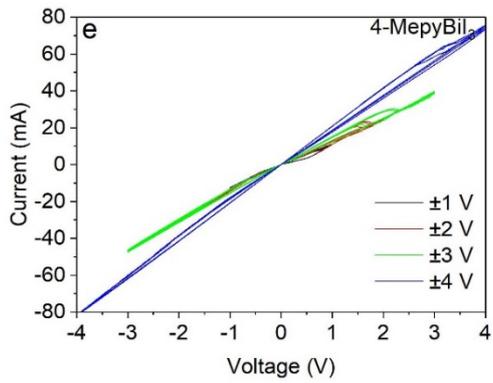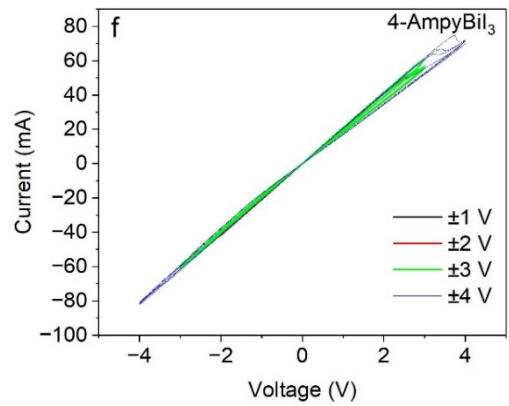

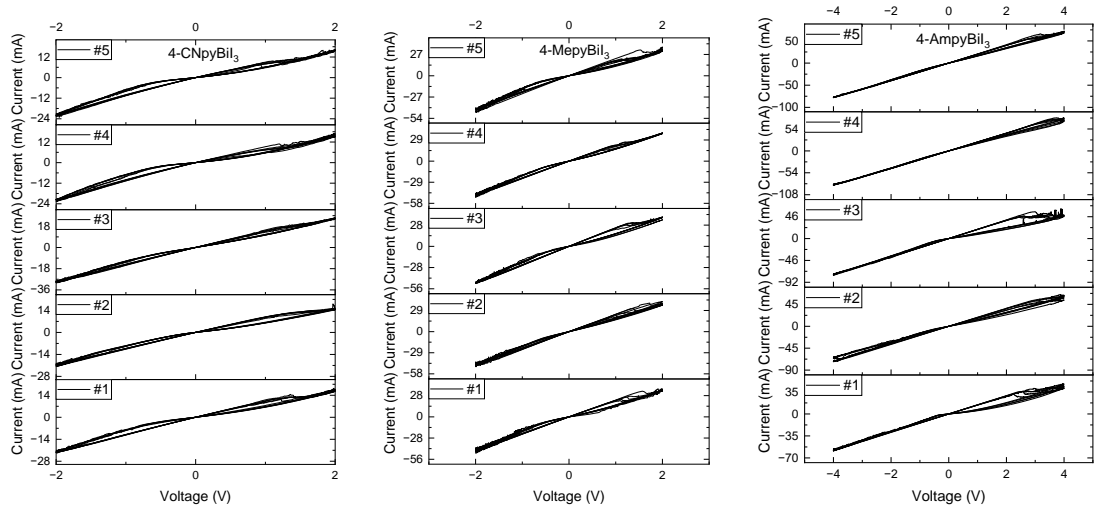

**Figure S4.** Scan rates a) Cu/4-CNpyBiI$_3$/ITO b) Cu/4-MepyBiI$_3$/ITO c) Cu/4-AmpyBiI$_3$/ITO. Voltage ranges d) Cu/4-CNpyBiI$_3$/ITO e) Cu/4-MepyBiI3/ITO f) Cu/4-AmpyBiI$_3$/ITO in different voltages range. Device-to-device reproducibility for 5 different batches of samples is also shown in bottom panel.

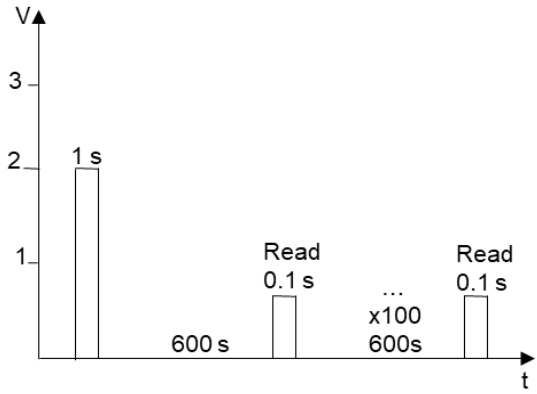

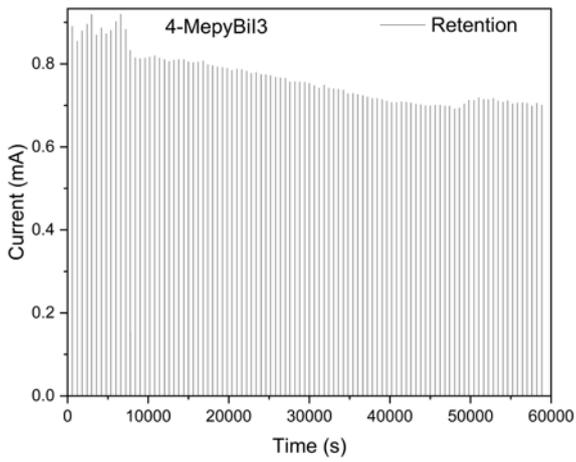

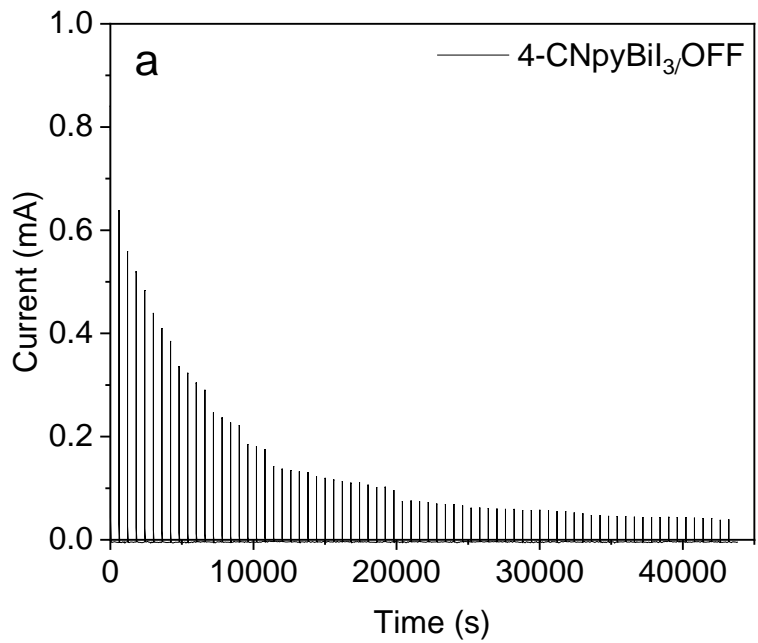

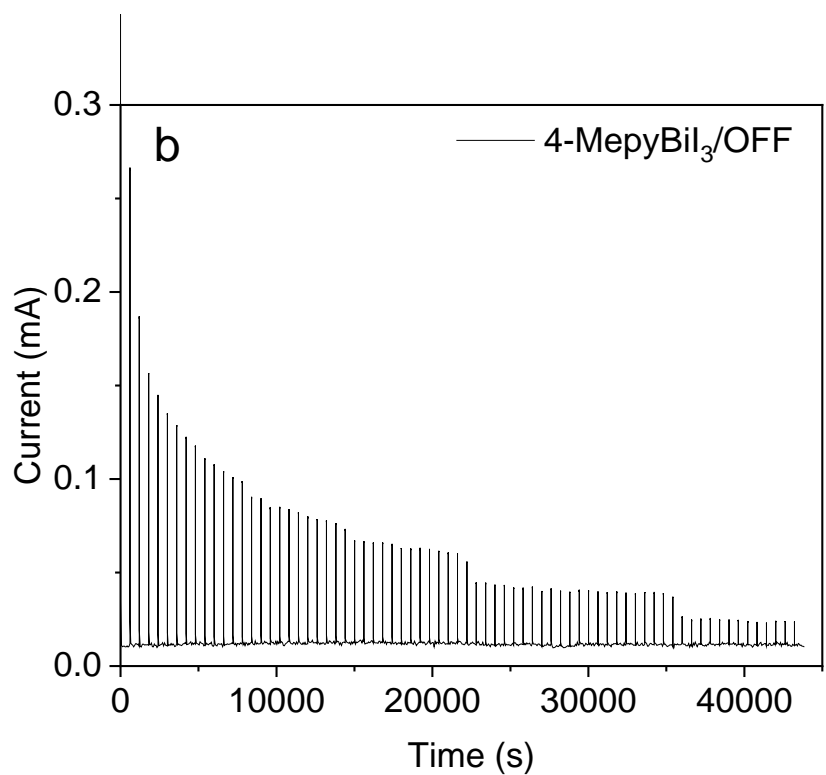

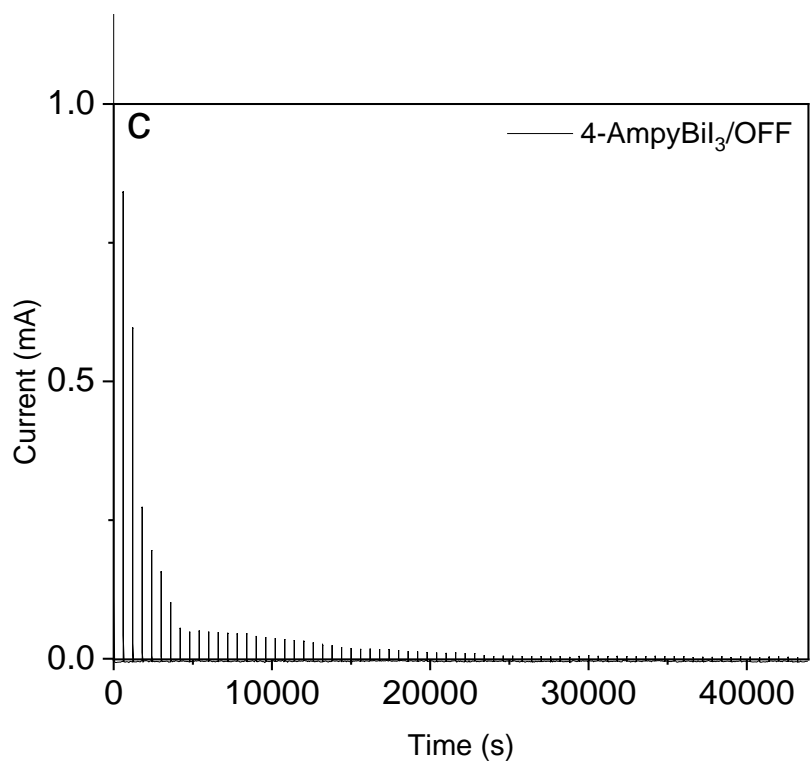

**Figure S5.** The Retention measurements of on and off states in 4-CNpyBiI$_3$ and 4-MepyBiI$_3$ at ±2 V as set and reset voltages. Retention measurements of the devices made of 4-AmpyBiI$_3$, ±4 V was chosen as the set and rest voltages and the read point was + 0.2 V. The width of all pulse was 0.1 s.

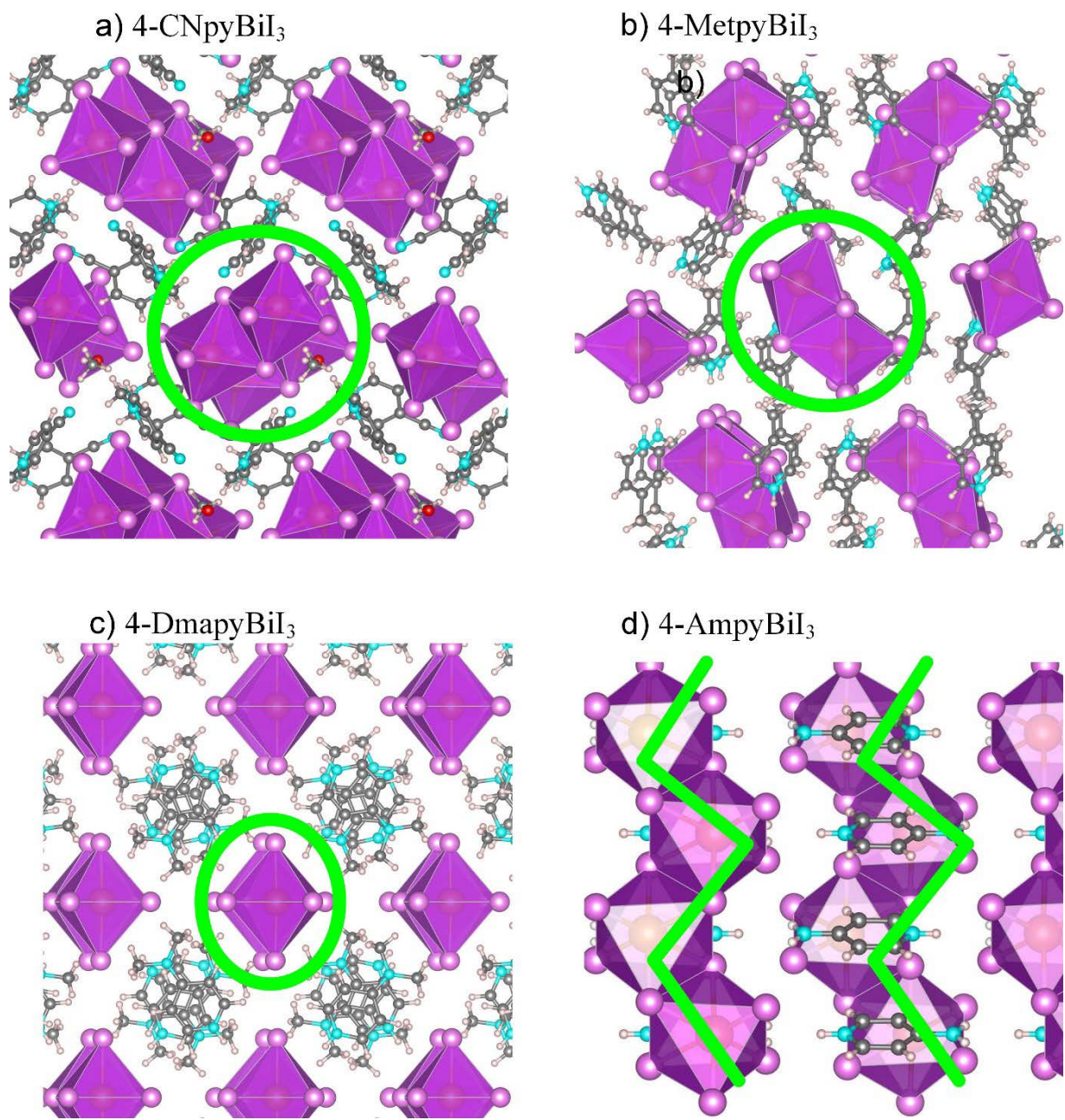

**Figure S6.** Representation of 0D a) 4-CNpyBiI$_3$ and b) 4-MepyBiI$_3$ c) 4-DmapyBiI$_3$ and 1D d) and 4-AmpyBiI$_3$ of ionic fragments of Bi-I and void shapes.

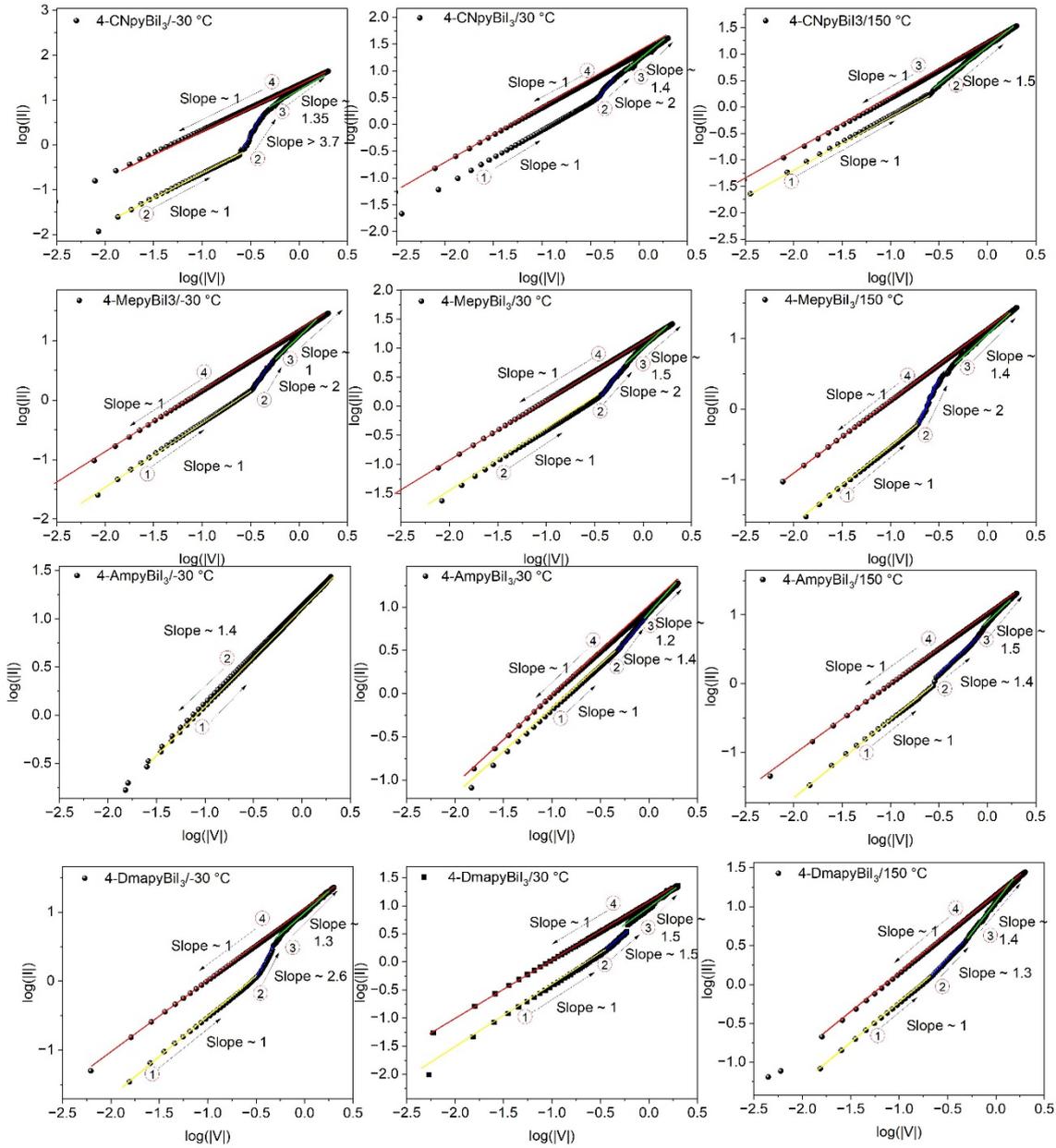

**Figure S7.** Double logarithmic scales of IV values of 4-CNpyBiI$_3$, 4-MepyBiI$_3$, 4-DmapyBiI$_3$ and 4-AmpyBiI$_3$ in SET process.

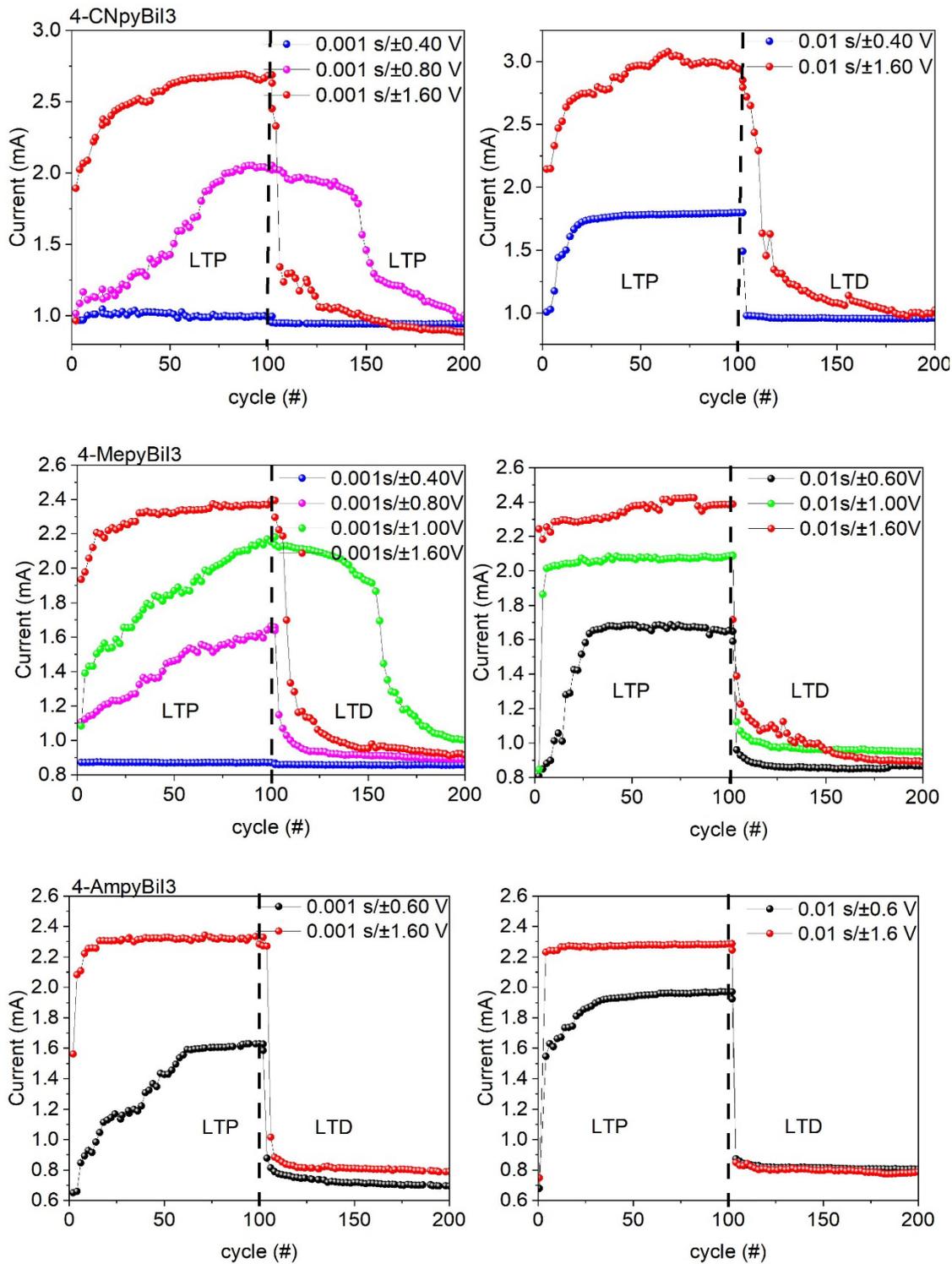

**Figure S8.** LTP and LTD for prepared devices with different pulse sequences with different width (0.001 s or 0.01 s) and amplitude 9between 0.4 V -1.6 V).

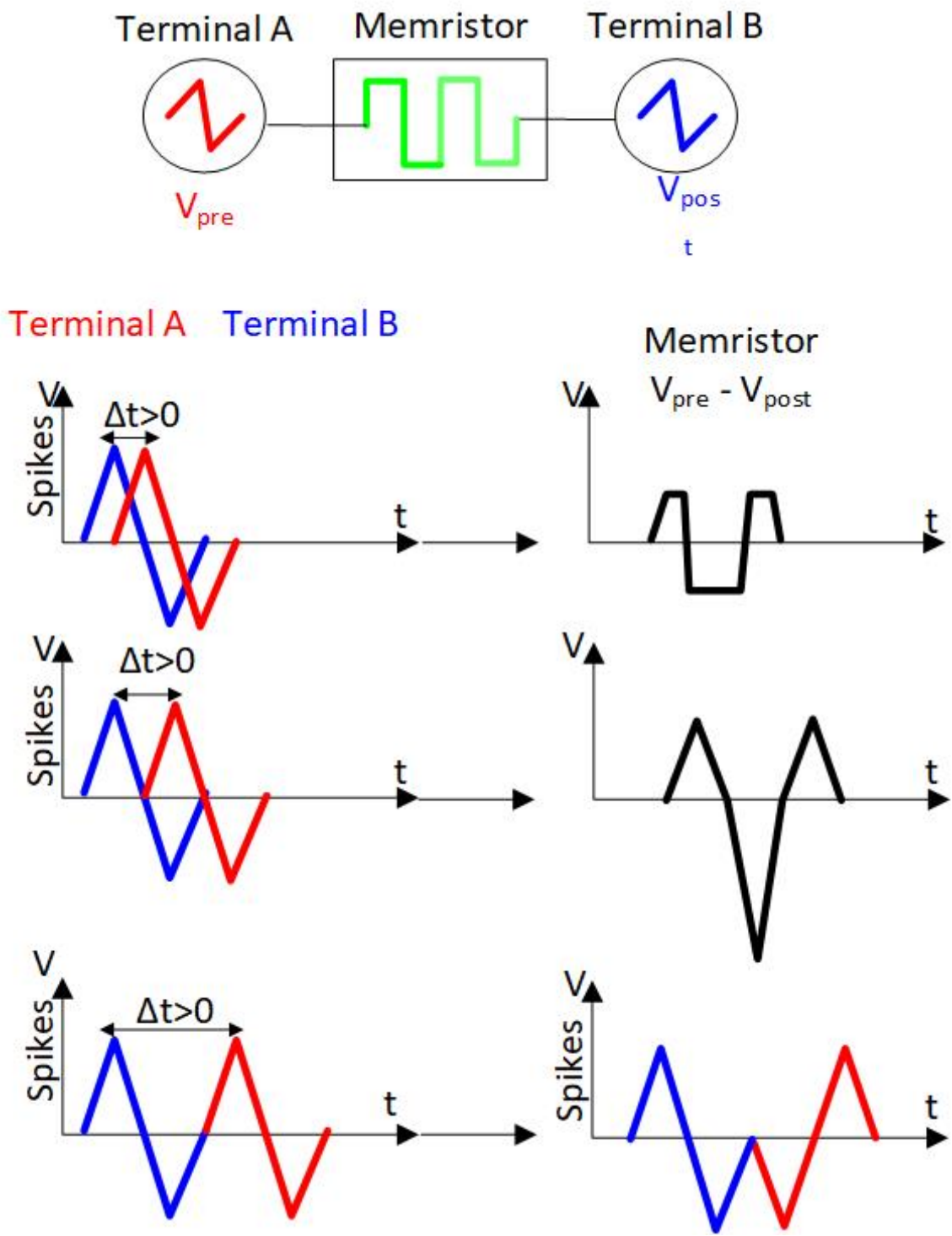

**Figure S9.** Illustration of final pulses shape on memristor devices, depends on time difference.

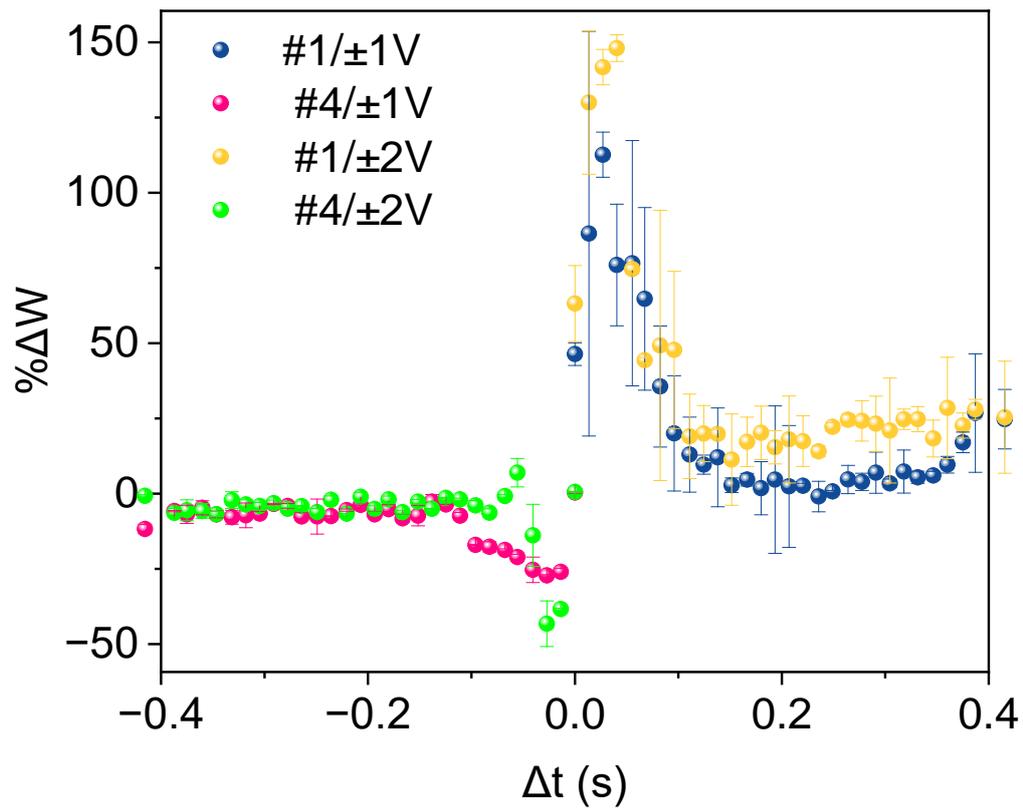

**Figure S10.** STDP of 4-CNpyBiI$_3$ with RT-pulse (#1 and #4) ; pulse amplitude to (±1 V) and pulse amplitude to (±2 V).

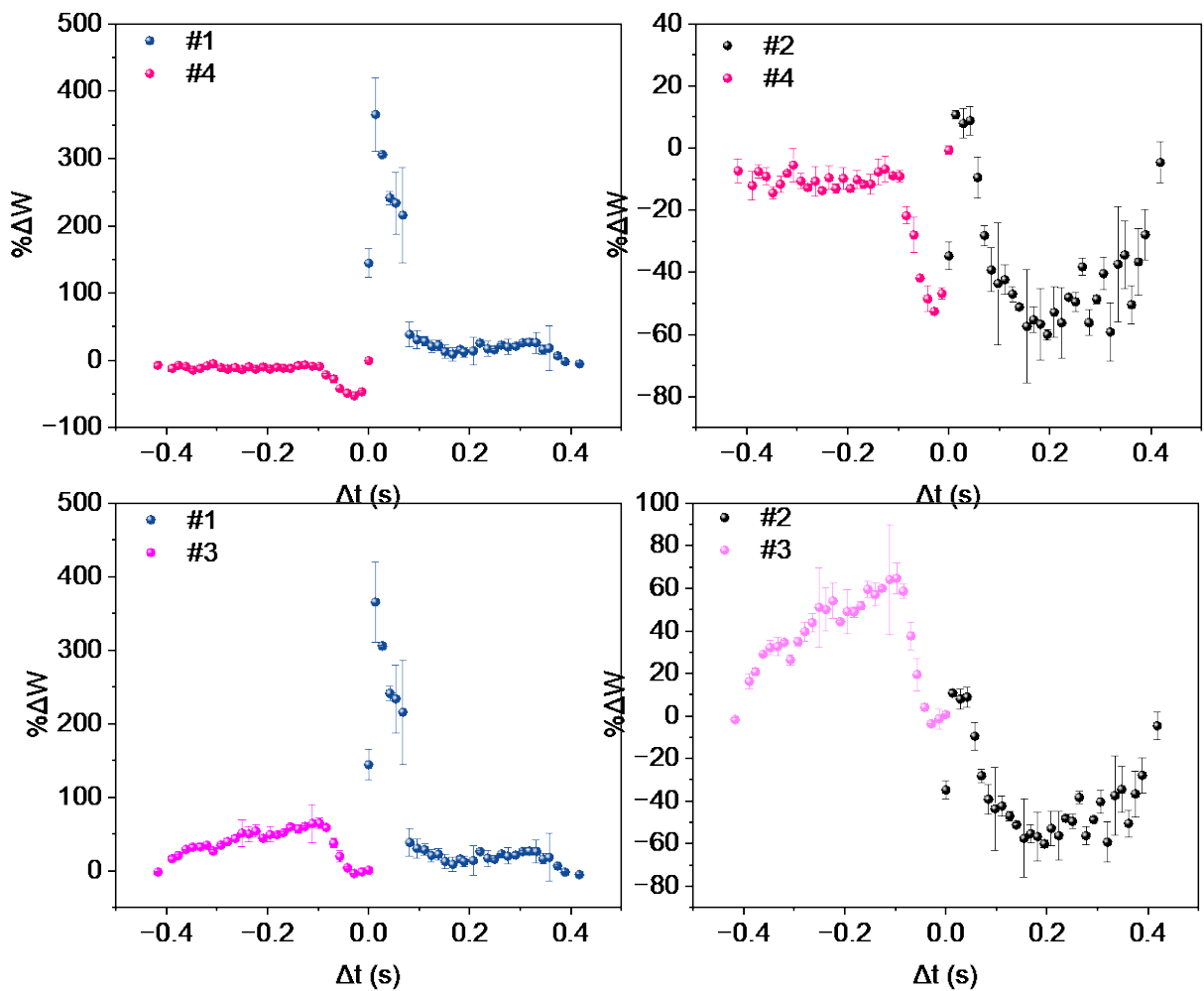

**Figure S11.** STDP of 4-MepyBiI$_3$ in different pulse polarities with ± 1 V amplitude.

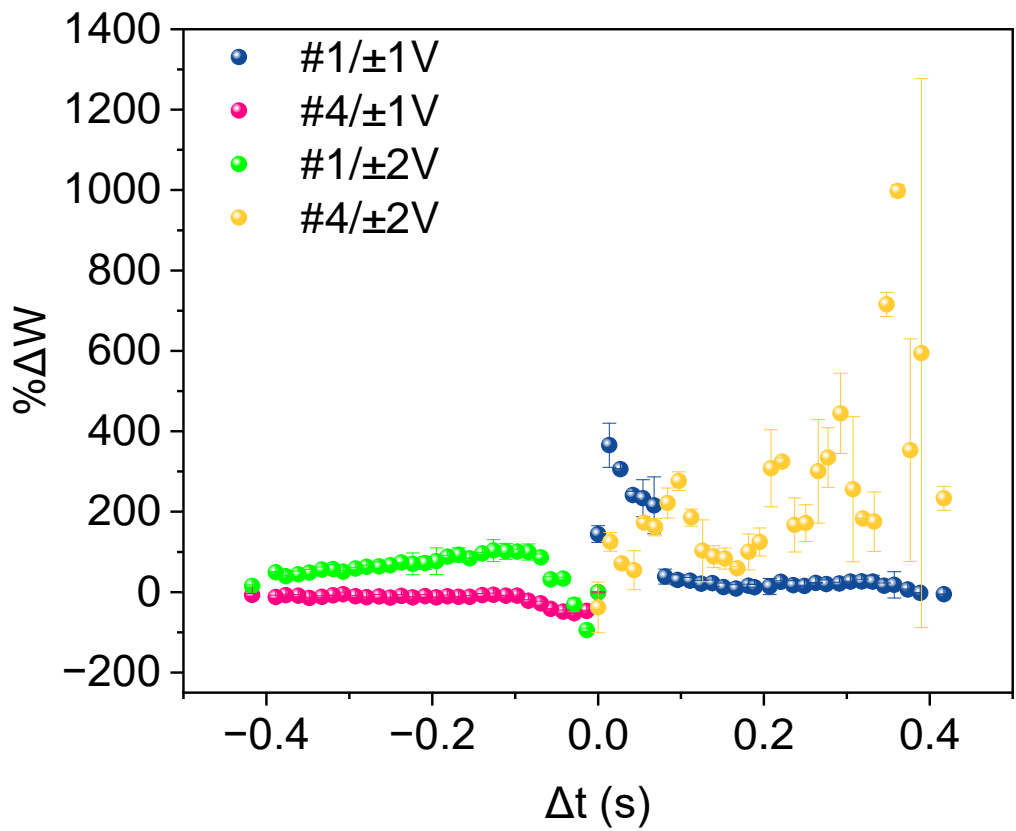

**Figure S12**. STDP of 4-MepyBiI$_3$ with RT-pulse (#1 and #4) ; pulse amplitude to (±1 V) and pulse amplitude to (±2 V).